\documentclass[reprint,amsmath,amssymb,aps,pre,showkeys,floatfix,longbibliography,superscriptaddress]{revtex4-2}

\usepackage{mathtools}
\usepackage{bm}
\usepackage{etoolbox}
\usepackage{ifthen}
\usepackage[dvipsnames]{xcolor}
\usepackage[normalem]{ulem}

\usepackage{mleftright}
\mleftright

\DeclareMathOperator{\sech}{sech}

\usepackage[pdfusetitle]{hyperref}
\hypersetup{
    breaklinks=true,
    colorlinks=true,
    citecolor=violet,
    linkcolor=MidnightBlue,
    urlcolor=orange,
}

\AtBeginEnvironment{thebibliography}{
    \interlinepenalty=10000
}

\begin{document}

\title{Continuum mean-field theory of spatiotemporal pattern formation in a three-state active Potts model}

\author{Yuta Tateyama}
\affiliation{Department of Physics, Chiba University, Chiba 263-8522, Japan}

\author{Songhao Yin}
\affiliation{Department of Physics, Chiba University, Chiba 263-8522, Japan}
\affiliation{Institute for Solid State Physics, University of Tokyo, Kashiwa, Chiba 277-8581, Japan}

\author{Hiroaki Ito}
\affiliation{Department of Physics, Chiba University, Chiba 263-8522, Japan}

\author{Hiroyuki Kitahata}
\email{kitahata@chiba-u.jp}
\affiliation{Department of Physics, Chiba University, Chiba 263-8522, Japan}

\author{Hiroshi Noguchi}
\email{noguchi@issp.u-tokyo.ac.jp}
\affiliation{Institute for Solid State Physics, University of Tokyo, Kashiwa, Chiba 277-8581, Japan}

\date{\today}

\begin{abstract}
We investigate the spatiotemporal dynamics of a three-state active Potts model with a cyclic transition bias.
Starting from microscopic Glauber-type dynamics, we derive macroscopic continuum field equations through a spatial mean-field approximation.
Linear stability and weakly nonlinear analyses reveal that increasing the coupling strength induces a Hopf bifurcation, near which the critical amplitude dynamics is described by the complex Ginzburg--Landau equation.
In the spatially homogeneous limit, the system exhibits limit-cycle oscillations and ordered states, with an intermediate regime of bistability bounded by a $\mathbb{Z}_3$-symmetric heteroclinic bifurcation.
In two dimensions, the system sustains stable rotating spiral waves.
Using perturbation theory based on one-dimensional traveling fronts, we semianalytically predict the curvature-dependent propagation velocity of circular fronts, as well as the rotational angular velocity and pitch of spiral waves.
Finally, we compare the continuum model with discrete Monte Carlo simulations to clarify the role of thermal fluctuations and evaluate the validity of the mean-field approximation.
\end{abstract}

\maketitle

\section{Introduction}
Spatiotemporal pattern formation in nonequilibrium open systems has been extensively studied across physics, chemistry, and biology~\cite{kuramoto1984chemical, cross1993pattern, cross2009pattern}.
In particular, spiral waves emerging in excitable or oscillatory media are widely observed in a variety of systems, ranging from cardiac arrhythmias~\cite{winfree1988time, winfree1989electrical, witkowski1998spatiotemporal} and aggregation processes of social amoebae~\cite{siegert1995spiral, hofer1995dictyostelium, palsson1996origin, lee1996competing} to chemical reaction systems such as the Belousov--Zhabotinsky reaction~\cite{muller1985structure, keener1986spiral, kitahata2011spontaneous}, catalytic surface reactions~\cite{bar1994spiral, von1998subsurface, kim2001controlling, bertram2003pattern, beta2004excitable, ertl2008reactions}, and precipitation processes~\cite{muller1982curiosities, volford2007pattern, pan2009precipitation, haudin2014spiral}.
In addition to these classical systems, nonequilibrium driving via nonreciprocal interactions has recently attracted significant attention for generating spontaneous oscillations and traveling waves~\cite{fruchart2021non}.

Fundamentally, breaking detailed balance via nonreciprocal interactions or directional state transitions drives the system out of equilibrium, sustaining probability currents, temporal oscillations, and traveling patterns.
For instance, in two-component nonreciprocal systems, traveling waves and persistent oscillations have been demonstrated in both discrete lattice models, such as the nonreciprocal Ising model~\cite{avni2025nonreciprocal, blom2025local, avni2025dynamical, arjun2026kinetic, blom2026dynamic}, and continuum field theories, including nonreciprocal Allen--Cahn~\cite{liu2023non}, Cahn--Hilliard~\cite{saha2020scalar, you2020nonreciprocity, frohoff2021suppression, frohoff2021localized, frohoff2023non, frohoff2023nonreciprocal, suchanek2023entropy, greve2024amplitude, saha2024phase, brauns2024nonreciprocal, greve2025coexistence}, and Swift--Hohenberg models~\cite{fruchart2021non, tateyama2024pattern, tateyama2026higher}.

In systems with three or more discrete states, transitions among internal states can form directed cycles, inherently generating a net circular probability flux.
This circular flux prevents the system from relaxing into a static equilibrium, thereby sustaining persistent directional transitions among states.
When combined with spatial coupling such as diffusion or nearest-neighbor interactions, such driven transition networks spontaneously organize into macroscopic spatiotemporal patterns, including traveling fronts and rotating spiral waves.
These cyclic models have been extensively applied to describe dynamic coexistence in ecosystems~\cite{may1975nonlinear, zhang2000global, petrovskii2001diffusive, dobramysl2018stochastic}, biodiversity in microbial populations~\cite{kerr2002local, bazeia2020breaking, serrao2021stabilizing}, pattern formation in chemically reacting or phase-separating mixtures~\cite{okuzono2001self, okuzono2003traveling, luo2023influence}, and dynamic encoding in neural networks~\cite{rabinovich2001dynamical}.
Theoretically, the emergence, wavelength selection, and stability of rotating spiral waves in spatial cyclic competition systems have been investigated using amplitude equations, such as the complex Ginzburg--Landau (CGL) equation~\cite{szczesny2013does, szczesny2014characterization}.

To incorporate spatial coupling and domain formation into such multistate active systems, the Potts model provides a prototypical lattice model.
Historically, the equilibrium Potts model~\cite{ashkin1943statistics, potts1952some, baxter1973potts, wu1982potts}, which generalizes the discrete spin states of the Ising model, has been widely applied to describe grain growth in materials science~\cite{holm2001computer}, as well as cell sorting~\cite{graner1992simulation} and tissue mechanics~\cite{rens2019energy, nemati2024cellular} in biological systems.
Incorporating nonequilibrium drive into the Potts model allows one to explore domain formation and traveling wave dynamics under broken detailed balance.
Recently, a three-state active Potts model with a cyclic transition bias was introduced using microscopic lattice Monte Carlo (MC) simulations~\cite{noguchi2024cycling, noguchi2024spatiotemporal}.
Subsequent studies extended the model to systems with four or more states, characterizing diverse wave modes and competing transition loops~\cite{noguchi2025spatiotemporal, noguchi2025dynamic, noguchi2026dynamics, noguchi2026spiral}.
Furthermore, recent studies have investigated domain coarsening dynamics and interfacial phenomena in driven active systems~\cite{blom2026mean, noguchi2026coarsening}, as well as applications to pattern formation in deformable membranes~\cite{noguchi2025spatiotemporal-membrane}.
The nonreciprocal Ising model corresponds to the four-state active vector Potts (clock) model~\cite{noguchi2025dynamic}.
In the active Potts model, the activity is implemented as a cyclic transition bias, whereas in the nonreciprocal Ising model, it arises from asymmetric interactions between paired spin species.

Coarse-graining of microscopic lattice models into continuum field theories provides analytical insight into pattern formation and self-organization mechanisms.
In self-propelled particle systems, for instance, coarse-graining of the microscopic Vicsek model~\cite{vicsek1995novel} leads to the hydrodynamic Toner--Tu equations~\cite{toner1995long, toner1998flocks, ramaswamy2010mechanics, marchetti2013hydrodynamics}.
In lattice systems, deriving continuum field theories such as the Allen--Cahn~\cite{glauber1963time, allen1979microscopic} and Cahn--Hilliard~\cite{cahn1958free, kawasaki1966diffusion} equations from microscopic master equations has long been established.
Recently, this approach has been extended to nonequilibrium systems by microscopically deriving nonreciprocal Allen--Cahn- and Cahn--Hilliard-type equations with zero, one, or two conservation laws from nonreciprocal Ising lattice models~\cite{blom2026dynamic}.
While nonreciprocal Ising models consider asymmetric coupling between distinct spin species, multistate active Potts models incorporate cyclic transitions among internal states within a single spin variable; however, a systematic microscopic derivation of their continuum field equations remains unexplored.

To bridge this gap, in this paper, we focus on the minimal three-state active Potts model with ferromagnetic nearest-neighbor interactions and a cyclic transition bias along the loop $s = 0 \rightarrow 1 \rightarrow 2 \rightarrow 0$.
The competition between the cyclic bias driving local directional transitions and the attractive spatial interactions sustains macroscopic oscillations and rotating spiral waves.
We construct a deterministic continuum field theory based on a spatial mean-field approximation, directly linking the microscopic transition rates to the macroscopic dynamics of traveling fronts and rotating spiral waves.
Formulating the macroscopic dynamics as continuum field equations enables us to analytically characterize the bifurcation structures and front propagation that are obscured by thermal fluctuations in discrete MC simulations.
Specifically, this continuum approach enables both a weakly nonlinear analysis near the Hopf onset (leading to the CGL equation) and a geometric eikonal theory for large-amplitude domain walls and spiral waves.

The remainder of this paper is organized as follows.
In Sec.~\ref{sec:model_continuum}, we introduce the three-state active Potts model and derive the deterministic continuum field equations via a spatial mean-field expansion.
Section~\ref{sec:theoretical_analysis} focuses on linear stability and weakly nonlinear analysis leading to the CGL equation, alongside an analysis of the $\mathbb{Z}_3$-symmetric heteroclinic bifurcation in the homogeneous limit.
In Sec.~\ref{sec:interfacial_and_spiral_dynamics}, we investigate large-amplitude interfacial dynamics, curvature-dependent domain-wall propagation, and the resulting rotating spiral waves based on an eikonal theory.
A quantitative comparison between the continuum predictions and lattice MC simulations is presented in Sec.~\ref{sec:mc_simulations} to evaluate the validity and limitations of the continuum approximation in the presence of thermal fluctuations and lattice discreteness.
Section~\ref{sec:discussion} discusses the physical implications and limitations of our continuum approach, and Sec.~\ref{sec:conclusion} concludes the paper.

\section{Active Potts Model and Continuum Formulation}\label{sec:model_continuum}
In this section, we introduce the microscopic three-state active Potts model on a two-dimensional (2D) lattice and derive its macroscopic continuum mean-field equations (CMFEs).
We then present representative numerical results obtained from the 2D CMFEs.

\subsection{Microscopic Hamiltonian and Cyclic Transition Bias}\label{sec:microscopic_model}
We consider a three-state active Potts model on a 2D lattice with lattice constant $\sigma$, where each site $i$ takes one of three discrete states $s \in \{0, 1, 2\}$.
The microscopic local potential $U_{i,s}$ at a site $i$ in state $s$ is defined as
\begin{equation}
    U_{i,s} \coloneqq - J \sum_{j \in \mathrm{nn}(i)} \delta_{s, s_j},
    \label{eq:micro_potential}
\end{equation}
where $J > 0$ is the ferromagnetic coupling strength, $\mathrm{nn}(i)$ represents the set of nearest neighbors of site $i$, excluding site $i$ itself, and $\delta_{s, s_j}$ is the Kronecker delta.
Based on this local potential, the total Hamiltonian of the system is defined as
\begin{equation}
    H(\bm{s}) \coloneqq \frac{1}{2} \sum_i U_{i, s_i},
    \label{eq:hamiltonian}
\end{equation}
where $\bm{s} = \{s_i\}$ represents the configuration of the states of all sites.
To introduce activity into the system, we apply a bias $h$ that drives transitions along a cycle in the state space.
In the following, all state indices and transitions $s \rightarrow s \pm 1$ are evaluated modulo 3.
Here, the forward bias is set to $h_{s, s+1} = h > 0$, while the backward bias is set to $h_{s, s-1} = -h < 0$.
A complete cycle $0 \rightarrow 1 \rightarrow 2 \rightarrow 0$ in the state space results in net work of $3 h$ performed on the system by the active drive, breaking detailed balance and generating persistent probability currents.
Assuming Glauber-type dynamics~\cite{glauber1963time}, the transition rate $w(s_i \rightarrow s'_i)$ for a local state transition at site $i$ is defined as
\begin{equation}
    w(s_i \rightarrow s'_i) \coloneqq \frac{1}{2\tau} \left[ 1 - \tanh\left( \frac{\beta}{2} \left( \Delta H - h_{s_i, s'_i} \right) \right) \right],
    \label{eq:Glauber_micro}
\end{equation}
where $\beta$ is the inverse temperature, $\tau$ is the characteristic transition time, and $\Delta H \coloneqq U_{i, s'_i} - U_{i, s_i}$ is the energy change associated with the transition.
For forward transitions $s \rightarrow s' = s+1$, the effective energy barrier is lowered by the positive bias $h$, thereby enhancing the transition rate, whereas backward transitions $s \rightarrow s' = s-1$ are suppressed by the opposing bias $-h$.

\subsection{Mean-Field and Spatial Continuum Approximation}\label{sec:continuum_derivation}
Let $\rho_{i,s}(t)$ be the local probability that site $i$ is in state $s$ at time $t$.
Neglecting microscopic fluctuations and correlations between neighboring sites, we perform a local mean-field approximation by replacing the microscopic state indicator $\delta_{s, s_j}$ with its local expectation value, i.e., the state probability $\rho_{j, s}(t)$.
Next, assuming that the state densities vary smoothly in space on scales much larger than the lattice constant $\sigma$, we take the spatial continuum limit $\sigma \to 0$, introducing continuous spatial coordinates $\bm{x}$.
Replacing the local probability $\rho_{i, s}(t)$ with a continuum field $\rho_s(\bm{x}, t)$ and expanding the nearest-neighbor sum up to the second order in spatial derivatives, we obtain
\begin{equation}
    \sum_{j \in \mathrm{nn}(i)} \rho_{j, s} \simeq z \rho_s(\bm{x}, t) + \frac{z}{4} \sigma^2 \nabla^2 \rho_s(\bm{x}, t),
\end{equation}
where $z$ is the coordination number of the lattice representing the number of nearest neighbors, such as $z = 4$ for a square lattice and $z = 6$ for a triangular lattice.
Through this continuum approximation, the continuum mean-field potential $U_s(\bm{x})$ experienced by a site in state $s$ is obtained as
\begin{equation}
    U_s(\bm{x}) = - z J \left( \rho_s + \frac{\sigma^2}{4} \nabla^2 \rho_s \right).
    \label{eq:mf_potential}
\end{equation}
Applying this spatial continuum mean-field approximation, we obtain the macroscopic CMFEs for the continuum density fields $\rho_s(\bm{x}, t)$:
\begin{align}
    \partial_t \rho_s &= w(s-1 \rightarrow s; \bm{x}) \rho_{s-1} + w(s+1 \rightarrow s; \bm{x}) \rho_{s+1} \notag\\
    &\quad - \left[ w(s \rightarrow s+1; \bm{x}) + w(s \rightarrow s-1; \bm{x}) \right] \rho_s,
    \label{eq:master_eq_contin}
\end{align}
where the continuous transition rate $w(s \rightarrow s'; \bm{x})$ at each position $\bm{x}$ is expressed in terms of the continuum mean-field potentials $U_s(\bm{x})$ as
\begin{align}
    &w(s \rightarrow s'; \bm{x}) \notag\\
    &= \frac{1}{2\tau} \left[ 1 - \tanh\left( \frac{\beta}{2} \left( U_{s'}(\bm{x}) - U_s(\bm{x}) - h_{s, s'} \right) \right) \right],
    \label{eq:transition_rate_contin}
\end{align}
and the total probability is conserved locally such that $\sum_{s=0}^{2} \rho_s(\bm{x}, t) = 1$.
For weak coupling (small $\beta z J$), the system settles into the homogeneous disordered state (trivial steady state), where the state probabilities are equipartitioned: $\rho_s(\bm{x}, t) = 1/3$ for all $s \in \{0, 1, 2\}$.
Conversely, for strong coupling (sufficiently large $\beta z J$), the system exhibits three homogeneous ordered states, where one of the three states dominates, e.g., $\rho_s \simeq 1$ and $\rho_{s'} \simeq 0$ for $s' \neq s$.

Unlike standard reaction-diffusion systems, the spatial Laplacian $\nabla^2 \rho_s$ appears inside the hyperbolic tangent function, meaning that the time evolution of each state is nonlinearly cross-coupled to the local densities and spatial Laplacians of all other states.
Furthermore, the active bias $h$ explicitly violates detailed balance, precluding the existence of a free-energy Lyapunov functional.
For convenience in the subsequent theoretical analysis, we introduce the dimensionless bias parameter $\zeta \coloneqq \tanh(\beta h / 2) \in [0, 1)$ to quantify the normalized drive strength.

\subsection{Spatiotemporal Patterns in the Continuum Model}\label{sec:pde_snapshots}
\begin{figure}[tbp]
    \centering
    \includegraphics[width=\columnwidth]{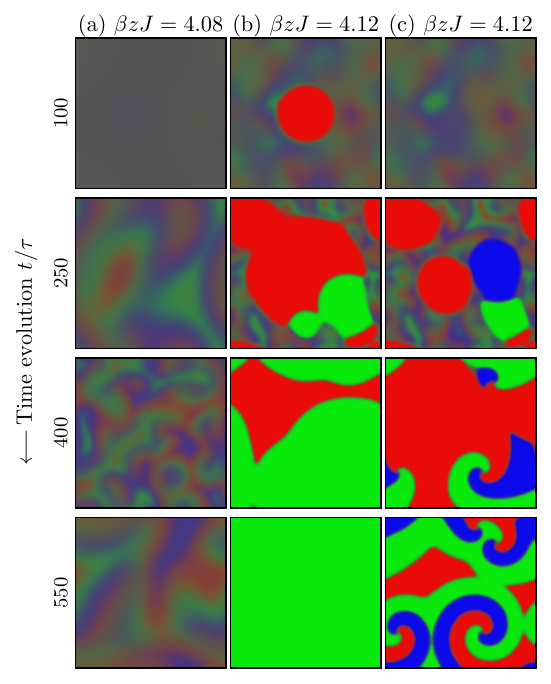}
    \caption{
        Time-series snapshots obtained by numerical simulations of the 2D CMFEs~\eqref{eq:mf_potential}--\eqref{eq:transition_rate_contin} at system size $L / \sigma = 128$ and bias parameter $\zeta = 0.5$.
        The RGB color channels represent the local densities of the three components as $(\rho_0, \rho_1, \rho_2)$ (red: $\rho_0$, green: $\rho_1$, and blue: $\rho_2$).
        In each column, the temporal evolution is shown from top to bottom at times $t / \tau = 100$, $250$, $400$, and $550$.
        (a) Small-amplitude spatiotemporal oscillations ($\beta z J = 4.08$) developing from white Gaussian noise around the homogeneous disordered state.
        (b) Domain growth leading to a uniform ordered state ($\beta z J = 4.12$) initiated from a circular nucleus ($s=0$, radius $R / \sigma = 10$).
        (c) A steadily rotating spiral wave sustained under the identical coupling strength ($\beta z J = 4.12$) initiated from white Gaussian noise without a nucleus.
        Columns (b) and (c) demonstrate the bistability between uniform ordering and rotating spiral waves governed by initial configurations.
    }
    \label{fig:pde_snapshots}
\end{figure}

To investigate the emerging spatiotemporal patterns, we perform 2D numerical simulations of the CMFEs~\eqref{eq:mf_potential}--\eqref{eq:transition_rate_contin} with system size $L/\sigma = 128$ and periodic boundary conditions.
Figure~\ref{fig:pde_snapshots} shows snapshots from these 2D CMFE simulations with a cyclic bias $\zeta = 0.5$ for two representative coupling strengths: $\beta z J = 4.08$ in (a), and $\beta z J = 4.12$ in (b) and (c).
Figure~\ref{fig:pde_snapshots}(a) shows small-amplitude spatiotemporal oscillations around the homogeneous disordered state just above the Hopf bifurcation point.
With increasing coupling strength, the system transitions to large-amplitude domain dynamics, exhibiting bistability where uniform ordered states and rotating spiral waves coexist depending on initial configurations, as shown in Figs.~\ref{fig:pde_snapshots}(b) and \ref{fig:pde_snapshots}(c).
When the initial configuration contains a circular nucleus of state $s = 0$ (shown in red with radius $R / \sigma = 10$) and relaxes without trapping topological defects, domains coarsen and expand to occupy the entire system, which eventually stabilizes into one of the three uniform ordered states, as shown in Fig.~\ref{fig:pde_snapshots}(b).
Conversely, when the system is initiated from white Gaussian noise without a nucleus at the same coupling strength, domain walls form a three-state junction with a nonzero topological winding, which acts as a protected core sustaining steadily rotating spiral waves, as shown in Fig.~\ref{fig:pde_snapshots}(c).
The underlying phase-space dynamics and bifurcation mechanisms governing these behaviors will be analyzed in the next section.

\section{Instability of the Disordered State and Bifurcation Structure}\label{sec:theoretical_analysis}
In this section, we elucidate how the instability of the homogeneous disordered state leads to spatiotemporal pattern formation in the CMFEs.
First, in Sec.~\ref{sec:linear_dispersion}, we linearize the governing equations around the disordered state to derive the dispersion relation and determine the Hopf bifurcation point, demonstrating that the absence of imaginary diffusion precludes the Benjamin--Feir instability near the onset.
Second, in Sec.~\ref{sec:cgl}, we perform a weakly nonlinear analysis to derive the CGL equation~\cite{kuramoto1984chemical} and identify the Bautin (generalized Hopf) bifurcation point~\cite{kuznetsov2023elements}, establishing that the weakly nonlinear oscillatory regime is confined to a narrow parameter window.
In Sec.~\ref{sec:homogeneous_limit_dynamics}, we investigate the phase-space dynamics of the spatially homogeneous system to elucidate the heteroclinic collapse of limit cycles, and demonstrate that rotating spiral waves in two dimensions are sustained by $\mathbb{Z}_3$-symmetric propagating domain walls connecting the coexisting ordered states around a topological defect core.

\subsection{Linear Stability and Dispersion Relation}\label{sec:linear_dispersion}
We linearize the CMFEs around the homogeneous disordered state $\rho_s = 1/3$ for all $s \in \{0, 1, 2\}$.
Letting $\delta\rho_s(\bm{x}, t) \coloneqq \rho_s(\bm{x}, t) - 1/3$ be the local perturbation of state $s$, the first-order expansion of the continuous transition rates~\eqref{eq:transition_rate_contin} yields the linearized evolution equation:
\begin{align}
    \tau \partial_t \delta\rho_s &= -\delta\rho_s + \frac{1+\zeta}{2} \delta\rho_{s-1} + \frac{1-\zeta}{2} \delta\rho_{s+1} \notag\\
    &\quad - \frac{\beta(1-\zeta^2)}{6} \left( 2 \delta U_s - \delta U_{s-1} - \delta U_{s+1} \right),
    \label{eq:master_eq_contin_linearized}
\end{align}
where the potential perturbation is given by $\delta U_s(\bm{x}, t) \coloneqq - z J [ \delta\rho_s + (\sigma^2 / 4) \nabla^2 \delta\rho_s ]$.
To determine the dispersion relation, we introduce the normal-mode ansatz $\delta\rho_s(\bm{x}, t) = \delta \tilde{\rho} e^{\lambda t} e^{i(\bm{q}\cdot\bm{x} + 2\pi s / 3)} + \mathrm{c.c.}$, where $\delta\tilde{\rho}$ is the complex amplitude, $\bm{q}$ is the wavevector with wavenumber $q = \|\bm{q}\|$, and $\mathrm{c.c.}$ denotes the complex conjugate.
Probability conservation imposes $\sum_s\delta\rho_s=0$, thereby eliminating the uniform mode.
The remaining perturbations therefore lie in the nontrivial $\mathbb{Z}_3$ representation sector, spanned by the conjugate cyclic phase factors $e^{2\pi i s/3}$ and $e^{-2\pi i s/3}$.
Substituting this perturbation into the linearized equation (for details of the derivation, see Appendix~\ref{app:linear_dispersion}), we derive the dispersion relation $\lambda(q)$ as
\begin{equation}
    \tau \lambda(q) = - \frac{3}{2} \left[ 1 - \frac{(1-\zeta^2) \beta z J}{3} \left( 1 - \frac{1}{4} \sigma^2 q^2 \right) \right] - i \frac{\sqrt{3}}{2} \zeta.
    \label{eq:dispersion_relation}
\end{equation}
From the marginal stability condition $\mathrm{Re}[\lambda(0)] = 0$ at wavenumber $q = 0$, which maximizes the growth rate, the Hopf bifurcation point $\beta z J_{\mathrm{Hopf}}$ and the corresponding oscillation frequency $\omega_0$ are determined as
\begin{equation}
    \beta z J_{\mathrm{Hopf}} = \frac{3}{1 - \zeta^2}, \quad
    \omega_0 = - \frac{\sqrt{3}}{2\tau} \zeta.
\end{equation}
The negative sign of $\omega_0$ reflects the clockwise rotation of the complex amplitude in the complex plane, corresponding to the forward cyclic transitions $0 \rightarrow 1 \rightarrow 2 \rightarrow 0$ driven by $\zeta > 0$.
In the vicinity of the bifurcation point, the spatial diffusion coefficient of the critical mode is obtained as $\gamma = 3 \sigma^2 / 8$, which is purely real and independent of the coordination number $z$.
Since this spatial diffusion coefficient has no imaginary component, the uniform oscillations are stable against the Benjamin--Feir (modulational) instability~\cite{kuramoto1984chemical} near the onset.

\subsection{Weakly Nonlinear Analysis}\label{sec:cgl}
To describe the universal amplitude dynamics near the Hopf bifurcation point, we perform a weakly nonlinear analysis of the CMFEs~\eqref{eq:mf_potential}--\eqref{eq:transition_rate_contin}.
We expand the local density fluctuations around the disordered state as $\delta\rho_s(\bm{x}, t) = A(\bm{x}, t) e^{2 \pi i s / 3} + \mathrm{c.c.}$, where $A(\bm{x}, t)$ is the complex amplitude of the critical mode.
Under the cyclic transition symmetry $s \rightarrow s + 1$, the discrete $\mathbb{Z}_3$ equivariance permits a nonresonant quadratic term $(A^*)^2$ in the amplitude equation alongside the cubic term.
By applying a near-identity transformation $A = B + \alpha (B^*)^2$ to absorb this quadratic contribution into the effective cubic nonlinearity, we reduce the evolution equation for the transformed amplitude $B(\bm{x}, t)$ to the standard CGL equation:
\begin{equation}
    \tau \partial_t B = (\mu + i \tau \omega_0) B + \gamma \nabla^2 B + g |B|^2 B + \mathcal{O}(|B|^4),
    \label{eq:cgl_equation}
\end{equation}
where $\mu \coloneqq (1 - \zeta^2) \beta z (J - J_{\mathrm{Hopf}}) / 2$ is the linear growth rate measuring the distance from the Hopf onset.
The detailed derivations of the expansion coefficients, the near-identity transformation, and the higher-order analysis up to the fifth order are presented in Appendices~\ref{app:amplitude_coefficients} and \ref{app:5th_order_nit}.

The supercritical or subcritical nature of the Hopf bifurcation is determined by the real part of the effective Landau coefficient, $\mathrm{Re}(g) = 81(3\zeta^2-1) / [8(1-\zeta^2)^2]$.
For $\zeta < 1/\sqrt{3} \simeq 0.577$, $\mathrm{Re}(g) < 0$ yields a supercritical Hopf bifurcation generating a stable limit cycle, whereas for $\zeta > 1/\sqrt{3}$, $\mathrm{Re}(g) > 0$ results in a subcritical Hopf bifurcation where the emerging limit cycle is unstable.
The threshold $\zeta_{\mathrm{c}} = 1/\sqrt{3}$ marks the Bautin (generalized Hopf) bifurcation point~\cite{kuznetsov2023elements} separating these two regimes.
This weakly nonlinear oscillatory regime is confined to a narrow parameter window near the Hopf onset, bounded by the $\mathbb{Z}_3$-symmetric heteroclinic collapse analyzed below.
Fully developed spiral waves lie outside this weakly nonlinear regime and are instead governed by propagating domain walls connecting the coexisting bulk ordered states.

\subsection{Dynamics and Bifurcations in the Spatially Homogeneous System}\label{sec:homogeneous_limit_dynamics}
\begin{figure*}[tbp]
    \centering
    \includegraphics[width=\textwidth]{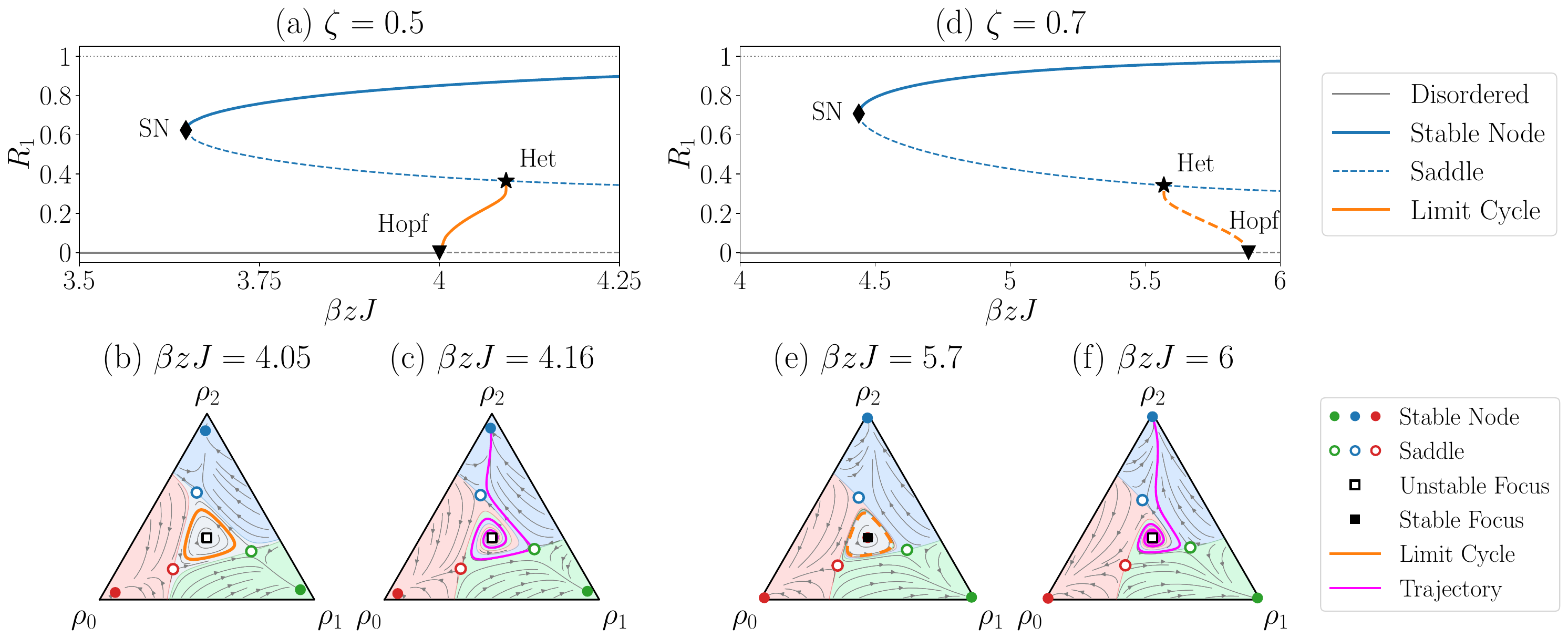}
    \caption{
        Bifurcation diagrams and phase portraits on the equilateral ternary simplex in the spatially homogeneous system of the CMFEs~\eqref{eq:mf_potential}--\eqref{eq:transition_rate_contin}.
        (a), (d) Bifurcation diagrams of the single-state order parameter $R_1$, defined in Eq.~\eqref{eq:homogeneous_order_parameter}, for (a) the supercritical regime ($\zeta=0.5$) and (d) the subcritical regime ($\zeta=0.7$).
        Blue curves denote the ordered states (solid for stable nodes and dashed for saddles), and orange curves show the limit cycles represented by their time-averaged order parameter $\overline{R}_1$ over one oscillation period (solid for stable and dashed for unstable).
        Downward triangles, diamonds, and stars mark the Hopf, $\mathbb{Z}_3$-symmetric saddle-node (SN), and $\mathbb{Z}_3$-symmetric heteroclinic (Het) bifurcation points, respectively.
        (b), (c), (e), (f) Phase portraits on the ternary simplex for (b), (c) the supercritical regime ($\zeta=0.5$) and (e), (f) the subcritical regime ($\zeta=0.7$).
        In (b) and (c), orange and magenta curves show a stable limit cycle ($\beta z J = 4.05$) and an outward escaping trajectory after the heteroclinic collapse ($\beta z J = 4.16$), respectively.
        In (e) and (f), orange and magenta curves show an unstable limit cycle ($\beta z J = 5.7$) and an outward escaping trajectory ($\beta z J = 6$), respectively.
        Solid colored circles, open colored circles, and square markers denote stable nodes, saddles, and the central focus (open for unstable and solid for stable), respectively.
        Gray streamlines, thin colored curves, and shaded regions show the phase-space flow, separatrices, and basins of attraction, respectively.
    }
    \label{fig:bifurcation_diagrams_combined}
\end{figure*}

To understand the global phase-space structure beyond the weakly nonlinear regime, we consider the spatially uniform state probabilities $(\rho_0(t), \rho_1(t), \rho_2(t))$ constrained by $\sum_{s=0}^2 \rho_s(t) = 1$, which define a two-variable ordinary differential equation (ODE) system on an equilateral ternary simplex.
To characterize the degree of single-state dominance, we introduce the order parameter
\begin{equation}
    R_1 \coloneqq \left| \sum_{s=0}^2 \rho_s(t) \exp\left( \frac{2\pi i s}{3} \right) \right|,
    \label{eq:homogeneous_order_parameter}
\end{equation}
which satisfies $R_1 \in [0, 1]$, taking $R_1 = 0$ for the homogeneous disordered state $(\rho_s = 1/3)$ and $R_1 = 1$ when a single state dominates.
Figure~\ref{fig:bifurcation_diagrams_combined} summarizes the bifurcation diagrams and representative phase portraits on the ternary simplex for both supercritical and subcritical regimes.
In the supercritical regime ($\zeta < 1/\sqrt{3}$), shown in Figs.~\ref{fig:bifurcation_diagrams_combined}(a)--(c) for $\zeta = 0.5$, when the coupling exceeds the Hopf bifurcation point $\beta z J_{\mathrm{Hopf}} = 4$, a stable limit cycle emerges around the central unstable focus and coexists with three stable ordered nodes created simultaneously via $\mathbb{Z}_3$-symmetric saddle-node bifurcations at $\beta z J_{\mathrm{SN}} \simeq 3.65$, as marked by the diamonds in Fig.~\ref{fig:bifurcation_diagrams_combined}(a).
Since $\beta z J_{\mathrm{SN}} < \beta z J_{\mathrm{Hopf}}$, the ordered states already exist at the Hopf onset, leading to multistability between the limit cycle and the ordered states.
With increasing coupling strength $\beta z J$, the limit cycle expands and deforms into a rounded triangular orbit, as shown in Fig.~\ref{fig:bifurcation_diagrams_combined}(b) for $\beta z J = 4.05$, until it collides with the three symmetric saddles at the $\mathbb{Z}_3$-symmetric heteroclinic bifurcation point $\beta z J_{\mathrm{het}} \simeq 4.092$.
Beyond this point ($\beta z J > \beta z J_{\mathrm{het}}$), the uniform limit cycle collapses, and trajectories starting near the disordered state escape outward toward one of the three stable ordered nodes, as shown in Fig.~\ref{fig:bifurcation_diagrams_combined}(c) for $\beta z J = 4.16$.
These stable nodes lie close to, but not strictly on, the simplex vertices ($R_1 \lesssim 1$), indicating that the ordered states retain minor fractions of the other two states.
For temporal waveforms of the limit cycle near the Hopf onset and the numerical scheme, see Fig.~\ref{fig:zeta-0_5-rho_time} in Supplemental Material (SM)~\cite{SM}, Sec.~\ref{sm:homogeneous_system}.

In the subcritical regime ($\zeta > 1 / \sqrt{3}$), shown in Figs.~\ref{fig:bifurcation_diagrams_combined}(d)--(f) for $\zeta = 0.7$, the system exhibits multistability between the disordered state and the three ordered states, with an unstable limit cycle enclosing the central disordered state.
As $\beta z J$ decreases, the unstable limit cycle expands, collides with the saddles, and disappears at the $\mathbb{Z}_3$-symmetric heteroclinic bifurcation point, while the coexistence persists until the stable nodes and saddles annihilate via $\mathbb{Z}_3$-symmetric saddle-node bifurcations at $\beta z J_{\mathrm{SN}} \simeq 4.44$.

\begin{figure}[tbp]
    \centering
    \includegraphics[width=\columnwidth]{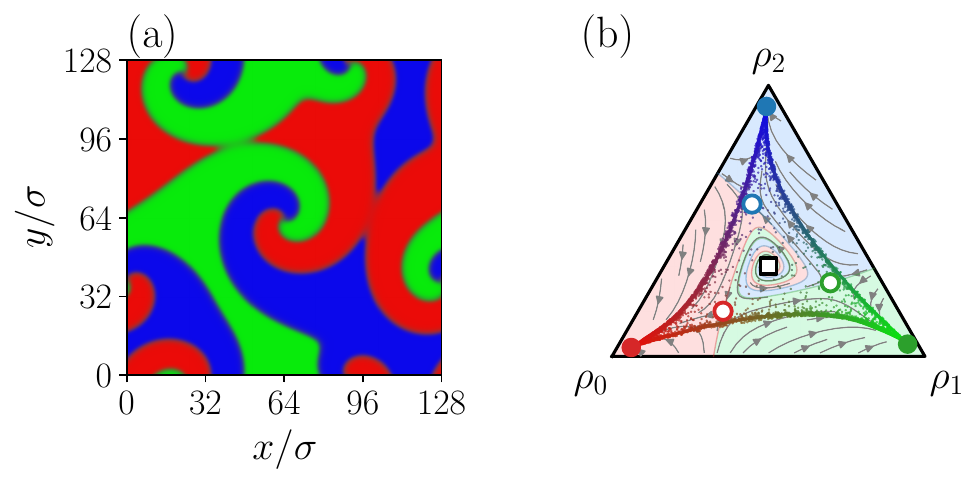}
    \caption{
        Rotating spiral wave and its projection onto the equilateral ternary simplex for $\zeta=0.5$ and $\beta z J = 4.16$.
        (a) Real-space snapshot obtained from 2D CMFE simulations~\eqref{eq:mf_potential}--\eqref{eq:transition_rate_contin}, using the same RGB color representation as in Fig.~\ref{fig:pde_snapshots}.
        (b) Scatter plot of the instantaneous local densities $(\rho_0(\bm{x}), \rho_1(\bm{x}), \rho_2(\bm{x}))$ evaluated at each spatial grid point from the snapshot in (a), projected onto the equilateral ternary simplex and overlaid on the phase portrait shown in Fig.~\ref{fig:bifurcation_diagrams_combined}(c).
    }
    \label{fig:simplex_spatial_scatter}
\end{figure}

Even in the parameter regime ($\beta z J > \beta z J_{\mathrm{het}}$) where uniform oscillations are extinguished by the heteroclinic collapse, the 2D spatial system maintains a steadily rotating spiral wave.
This is because different ordered states can coexist spatially, separated by domain walls that propagate unidirectionally under the cyclic bias.
At the junction where three such domains meet, the local densities balance at the disordered state $\rho_s = 1/3$ for all $s \in \{0, 1, 2\}$ with $R_1 = 0$.
Along a closed path enclosing this junction, the phase $\arg\left[ \sum_{s=0}^2 \rho_s e^{2\pi i s/3} \right]$ winds by $2\pi$, forming a phase singularity that prevents the junction from disappearing and sustains persistent rotation as a topologically protected spiral core.
To elucidate how this spatiotemporal pattern relates to the underlying phase-space structure, Fig.~\ref{fig:simplex_spatial_scatter}(a) displays a real-space snapshot of the rotating spiral wave obtained from 2D CMFE simulations for $\zeta=0.5$ and $\beta z J = 4.16$, while Fig.~\ref{fig:simplex_spatial_scatter}(b) shows the corresponding local densities projected onto the equilateral ternary simplex.
In Fig.~\ref{fig:simplex_spatial_scatter}(b), the projected local densities concentrate near the three stable ordered nodes corresponding to the domain interiors, while the propagating domain walls form the inwardly curved triangular contour connecting them.
This projection demonstrates that the rotating spiral wave is sustained by traveling domain walls connecting the coexisting ordered states, rather than by a limit-cycle oscillation of the spatially homogeneous system.
Consequently, determining the rotational angular velocity and pitch of spiral waves requires shifting from homogeneous oscillatory dynamics to the propagation properties of one-dimensional (1D) traveling fronts connecting coexisting ordered states, which we analyze in the next section.

\section{Large-Amplitude Interfacial Dynamics and Spiral Waves}\label{sec:interfacial_and_spiral_dynamics}
In this section, we investigate the interfacial dynamics and large-amplitude rotating spiral waves.
In Sec.~\ref{sec:traveling_front}, we formulate 1D steady traveling front solutions as a boundary value problem (BVP) to evaluate the intrinsic flat propagation velocity $V_{\mathrm{flat}}$.
In Sec.~\ref{sec:circular_front}, we extend the analysis to two dimensions by simulating circular domain fronts in the 2D CMFEs, validating the eikonal equation and deriving the effective diffusion coefficient $D_{\mathrm{eff}}$ via first-order perturbation theory.
Finally, in Sec.~\ref{sec:spiral_waves}, we apply Keener's eikonal theory~\cite{keener1986geometrical, tyson1988singular} to theoretically predict the rotational angular velocity and pitch of spiral waves from $V_{\mathrm{flat}}$ and $D_{\mathrm{eff}}$, comparing them with 2D CMFE simulations.

\subsection{One-Dimensional Steady Traveling Front Dynamics}\label{sec:traveling_front}
As demonstrated in the preceding section, large-amplitude spiral waves are sustained by $\mathbb{Z}_3$-symmetric propagating domain walls connecting the three coexisting bulk ordered states around a central topological defect, rather than by small-amplitude phase oscillations.
In two dimensions, the normal velocity of such a domain wall depends on the intrinsic propagation velocity of a flat interface, $V_{\mathrm{flat}}$, with the local curvature correction, serving as the essential inputs for the geometric spiral theory developed below.
As the first step, we determine the intrinsic flat propagation velocity $V_{\mathrm{flat}}$ by formulating 1D steady traveling fronts as a nonlinear BVP.
In the far field of a spiral arm, the interface curvature vanishes and the local wave profile asymptotically approaches that of a flat front, making this 1D solution the fundamental reference for the pattern dynamics.

To find a front propagating steadily with a constant velocity $V$, we introduce the comoving coordinate $\xi \coloneqq x - V t$.
Substituting this coordinate into the CMFEs~\eqref{eq:mf_potential}--\eqref{eq:transition_rate_contin}, we obtain the ODE system governing the traveling front:
\begin{align}
    - V \frac{d \rho_s}{d \xi} &= w(s-1 \rightarrow s; \xi) \rho_{s-1} + w(s+1 \rightarrow s; \xi) \rho_{s+1} \notag\\
    &\quad - \left[ w(s \rightarrow s+1; \xi) + w(s \rightarrow s-1; \xi) \right] \rho_s.
    \label{eq:traveling_front_ode}
\end{align}
We solve this 1D BVP numerically by imposing boundary conditions connecting two adjacent ordered states at $\xi \to \pm\infty$ (see SM~\cite{SM}, Sec.~\ref{sm:bvp_numerical_scheme}, for numerical details).

\begin{figure*}[tbp]
    \centering
    \includegraphics[width=\textwidth]{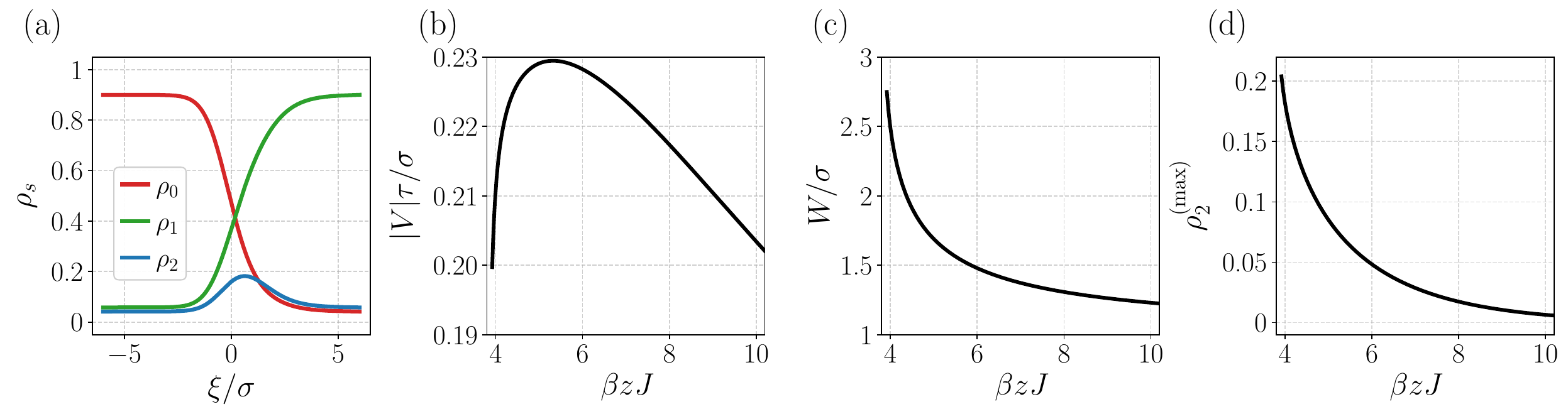}
    \caption{
        Characteristics of the 1D steady traveling front obtained from the ODE system~\eqref{eq:traveling_front_ode} at bias parameter $\zeta = 0.5$.
        (a) Spatial profiles of the three state densities $\rho_s$ across the front for $\beta z J = 4$, where the intermediate state $s = 2$ forms a localized peak in $\rho_2$ around the interface.
        (b)--(d) Dependence on the coupling strength $\beta z J$ of (b) the dimensionless front propagation velocity $|V|\tau/\sigma$, (c) the dimensionless interface width $W/\sigma$, and (d) the maximum occupancy of state $s = 2$, $\rho_2^{(\mathrm{max})}$.
    }
    \label{fig:bvp_summary}
\end{figure*}

Figure~\ref{fig:bvp_summary}(a) shows the steady traveling front profile obtained from Eq.~\eqref{eq:traveling_front_ode} for $\zeta = 0.5$.
Driven by the active bias, the third state $s = 2$ is induced around the interface, forming a distinct peak in $\rho_2$.
This implies that although the front macroscopically separates ordered states $s = 0$ and $s = 1$, the third state also participates in the local interface structure.

Figure~\ref{fig:bvp_summary}(b) shows the dependence of the dimensionless front velocity $|V|\tau/\sigma$ on the coupling strength $\beta z J$, which exhibits a nonmonotonic behavior reaching a maximum of $|V|\tau/\sigma \simeq 0.229$ around $\beta z J \simeq 5.32$.
Figures~\ref{fig:bvp_summary}(c) and \ref{fig:bvp_summary}(d) display the corresponding dimensionless interface width $W/\sigma$ (defined in SM~\cite{SM}, Sec.~\ref{sm:1d_pde_scheme}) and the maximum occupancy of the intermediate state $\rho_2^{(\mathrm{max})}$, respectively.
With increasing coupling strength $\beta z J$, both the dimensionless interface width $W/\sigma$ and the maximum occupancy $\rho_2^{(\mathrm{max})}$ decrease monotonically.
Direct numerical simulations of the 1D CMFEs confirm that these traveling front solutions propagate steadily without exhibiting any instability, as detailed in SM~\cite{SM}, Sec.~\ref{sm:pde_verification}.

\begin{figure}[tbp]
    \centering
    \includegraphics[width=\columnwidth]{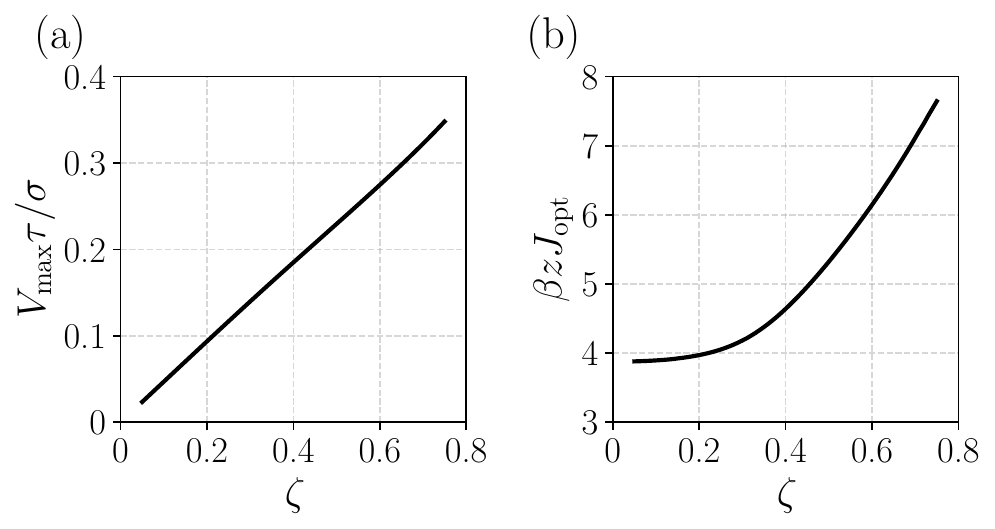}
    \caption{
        Dependence on the bias parameter $\zeta$ of (a) the maximum dimensionless propagation velocity $V_{\mathrm{max}}\tau/\sigma$ and (b) the optimal coupling strength $\beta z J_{\mathrm{opt}}$ yielding the maximum velocity, obtained from the ODE system~\eqref{eq:traveling_front_ode}.
    }
    \label{fig:zeta_vs_argmax-zJ_Vmax}
\end{figure}

Figure~\ref{fig:zeta_vs_argmax-zJ_Vmax} shows the dependence on the bias parameter $\zeta$ of the maximum dimensionless velocity $V_{\mathrm{max}}\tau/\sigma$ and the corresponding optimal coupling strength $\beta z J_{\mathrm{opt}}$ that maximizes the front velocity for each $\zeta$.
Numerically, the maximum dimensionless velocity $V_{\mathrm{max}}\tau/\sigma$ is found to be proportional to $\zeta$, while $\beta z J_{\mathrm{opt}}$ increases monotonically with $\zeta$.

\subsection{Two-Dimensional Curved Interfaces and Curvature-Dependent Front Propagation}\label{sec:circular_front}
To quantify the geometric motion of curved interfaces, we evaluate the normal propagation velocity of circular domain fronts.
On a 2D grid, we set a circular domain of state $s$ with an initial radius $R(0) = R_0$ in a uniform background of state $s-1$ or state $s+1$.
Due to the cyclic transition preference $s-1 \rightarrow s$, the domain interior of state $s$ invades outward into the background of state $s-1$, forming a front that is convex with respect to the forward invasion direction.
Conversely, the background of state $s+1$ invades inward into the domain interior of state $s$, forming a front that is concave with respect to the forward invasion direction.
For sufficiently large initial radii $R_0$, the convex domain expands outward with $dR/dt > 0$, whereas the concave domain collapses inward with $dR/dt < 0$.
For these curved front configurations, the domain wall propagates radially with the normal propagation velocity $V_{\mathrm{n}}$ in the forward invasion direction, evaluated as $V_{\mathrm{n}} \coloneqq dR/dt$ for convex fronts and $V_{\mathrm{n}} \coloneqq -dR/dt$ for concave fronts.
In the small-curvature regime, this normal velocity obeys the eikonal equation~\cite{keener1986geometrical, tyson1988singular}:
\begin{equation}
    V_{\mathrm{n}} = V_{\mathrm{flat}} - D_{\mathrm{eff}} \kappa,
    \label{eq:eikonal}
\end{equation}
where $V_{\mathrm{flat}} > 0$ is the velocity of the flat traveling front obtained from the 1D BVP in Sec.~\ref{sec:traveling_front}, $D_{\mathrm{eff}}$ is the effective diffusion coefficient, and the signed curvature is defined as $\kappa \coloneqq 1/R > 0$ for convex fronts and $\kappa \coloneqq -1/R < 0$ for concave fronts.
Note that $D_{\mathrm{eff}}$ represents the geometric interface diffusion coefficient determined by the front profile, distinct from the linear diffusion coefficient $\gamma$ of the critical mode in Sec.~\ref{sec:cgl}.

\begin{figure}[tbp]
    \centering
    \includegraphics[width=\columnwidth]{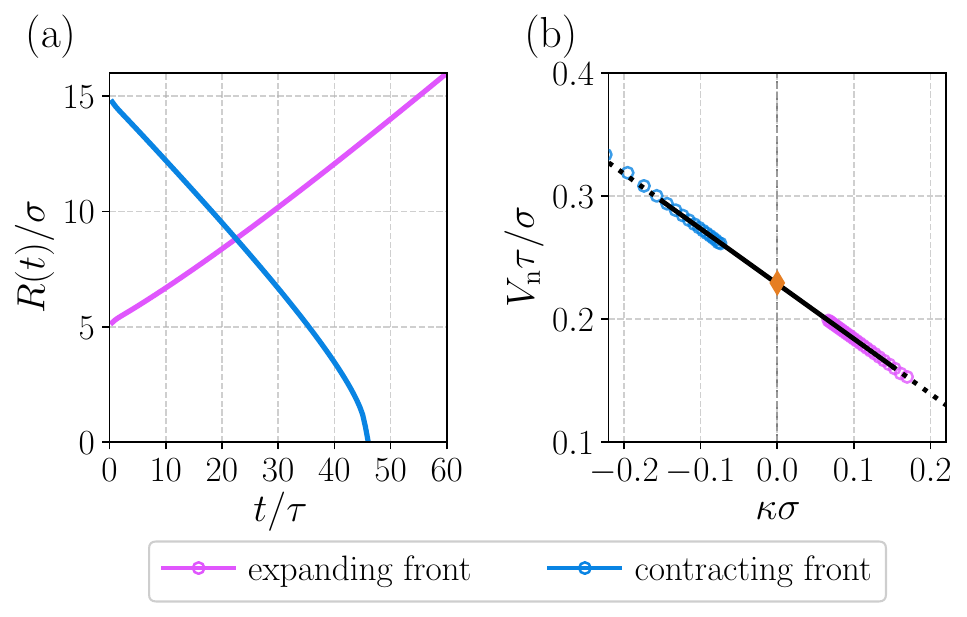}
    \caption{
        Curvature-dependent front propagation of circular domains in 2D CMFE simulations~\eqref{eq:mf_potential}--\eqref{eq:transition_rate_contin} for $\beta z J = 5.5$ at bias parameter $\zeta = 0.5$.
        (a) Time evolution of the dimensionless domain radius $R(t)/\sigma$ for expanding and contracting circular domains.
        (b) Dimensionless normal interface velocity $V_{\mathrm{n}}\tau/\sigma$ as a function of the dimensionless signed curvature $\kappa\sigma$, where $\kappa\sigma > 0$ corresponds to expansion and $\kappa\sigma < 0$ to contraction.
        Magenta and blue open circles denote 2D CMFE simulation data for expanding and contracting fronts, respectively.
        The black solid line represents the linear fit to the data within the small-curvature regime $|\kappa|\sigma \leq 0.15$, with the black dotted lines showing its extrapolation.
        The orange diamond indicates the dimensionless flat-interface propagation velocity $V_{\mathrm{flat}}\tau/\sigma$ obtained from the 1D BVP analysis in Sec.~\ref{sec:traveling_front}.
    }
    \label{fig:circular_front_individual}
\end{figure}

Tracking the domain radius $R(t)$ over time, we calculate the normal velocity $V_{\mathrm{n}}$ as a function of the signed curvature $\kappa$ at each time step.
Figure~\ref{fig:circular_front_individual} shows the analysis results of the curvature-dependent front propagation for a typical coupling strength $\beta z J = 5.5$ at bias parameter $\zeta = 0.5$.
As shown in Fig.~\ref{fig:circular_front_individual}(a), the expanding domain grows almost steadily, whereas the contracting domain accelerates as its radius shrinks, exhibiting a rapid collapse toward $R \to +0$.
To eliminate finite-size effects arising from the accelerated collapse of small contracting domains, the linear eikonal regression in Fig.~\ref{fig:circular_front_individual}(b) is restricted to the small-curvature regime $|\kappa| \sigma \leq 0.15$, yielding the linear fit $V_{\mathrm{n}}\tau/\sigma \simeq 0.229 - 0.45 \kappa\sigma$, shown by the black solid line.
From a comparison of this fit with the eikonal equation~\eqref{eq:eikonal}, the negative slope yields the dimensionless effective diffusion coefficient $D_{\mathrm{eff}}\tau/\sigma^2 \simeq 0.45$.
Furthermore, the intercept at $\kappa = 0$ provides the estimated dimensionless flat-interface propagation velocity $V_{\mathrm{flat}}\tau/\sigma \simeq 0.229$, which quantitatively agrees with the 1D BVP velocity plotted as the orange diamond.
Extrapolating this linear eikonal equation to $V_{\mathrm{n}} = 0$ provides an estimate for the critical nucleus radius $R_{\mathrm{c}} \coloneqq D_{\mathrm{eff}}/V_{\mathrm{flat}}$, below which the velocity reduction $D_{\mathrm{eff}}\kappa$ due to curvature overcomes the propagation velocity $V_{\mathrm{flat}}$ of the flat interface and causes spontaneous domain shrinkage even in the convex configuration, as shown in Fig.~\ref{fig:sm_critical_nucleus} in SM~\cite{SM}, Sec.~\ref{sm:circular_front_scheme}.

To determine the effective diffusion coefficient $D_{\mathrm{eff}}$ theoretically, we apply the first-order perturbation theory around the flat traveling front.
In the comoving coordinate $\xi$ along the normal to a circular front, we expand the front profile in powers of the local curvature $\kappa$:
\begin{equation}
    \bm{\rho}(\xi) = \bm{\rho}^{(0)}(\xi) + \kappa \bm{\rho}^{(1)}(\xi) + \mathcal{O}(\kappa^2),
    \label{eq:front_curvature_expansion}
\end{equation}
where $\bm{\rho}^{(0)}(\xi)$ is the flat 1D traveling front solution of Eq.~\eqref{eq:traveling_front_ode}, and $\bm{\rho}^{(1)}(\xi)$ represents the perturbation of the front profile induced by curvature.
Linearizing the CMFEs to first order in $\kappa$ with Eq.~\eqref{eq:front_curvature_expansion} and the eikonal equation~\eqref{eq:eikonal} yields the perturbation equation:
\begin{equation}
    \mathcal{J} \bm{\rho}^{(1)} = - D_{\mathrm{eff}} \partial_\xi \bm{\rho}^{(0)} + \bm{N}(\bm{\rho}^{(0)}),
    \label{eq:linear_inhomogeneous_ODE}
\end{equation}
where $\mathcal{J}$ is the linearized Jacobian operator around the flat front profile $\bm{\rho}^{(0)}(\xi)$, and $\bm{N}(\bm{\rho}^{(0)})$ is the curvature-induced term (see Appendix~\ref{app:curvature_perturbation_theory} for explicit algebraic expressions).
Due to translational symmetry, the Fredholm alternative requires the inhomogeneous vector on the right-hand side of Eq.~\eqref{eq:linear_inhomogeneous_ODE} to be orthogonal to the adjoint null-vector $\bm{\phi}^\dagger(\xi)$ of $\mathcal{J}$, which satisfies $\mathcal{J}^\dagger \bm{\phi}^\dagger = 0$ (see Appendix~\ref{app:curvature_perturbation_theory} for details).
Projecting Eq.~\eqref{eq:linear_inhomogeneous_ODE} onto $\bm{\phi}^\dagger(\xi)$ yields the integral expression for the effective diffusion coefficient $D_{\mathrm{eff}}$:
\begin{equation}
    D_{\mathrm{eff}} = \frac{\int_{-\infty}^{\infty} \bm{\phi}^\dagger(\xi) \cdot \bm{N}(\bm{\rho}^{(0)}) d\xi}{\int_{-\infty}^{\infty} \bm{\phi}^\dagger(\xi) \cdot \partial_\xi \bm{\rho}^{(0)} d\xi}.
    \label{eq:deff_integral_ratio}
\end{equation}

Figure~\ref{fig:circular_front_sweep_summary} shows the dependence of the front dynamics on the coupling strength $\beta z J$.
As shown in Fig.~\ref{fig:circular_front_sweep_summary}(a), the flat-interface propagation velocity $V_{\mathrm{flat}}\tau/\sigma$ extracted from the 2D CMFE simulations (blue circles) matches the 1D BVP solution (black solid line) over the entire range of $\beta z J$.
Furthermore, as shown in Fig.~\ref{fig:circular_front_sweep_summary}(b), the theoretical predictions of $D_{\mathrm{eff}}\tau/\sigma^2$ evaluated via Eq.~\eqref{eq:deff_integral_ratio} (orange solid line) quantitatively reproduce the coupling-strength dependence observed in the simulations (blue circles).
These results demonstrate that the 2D curved interface dynamics in this model is consistently described by the eikonal theory based on the 1D front solution.

\subsection{Rotating Spiral Waves and Finite-Core Eikonal Theory}\label{sec:spiral_waves}
\begin{figure}[tbp]
    \centering
    \includegraphics[width=\columnwidth]{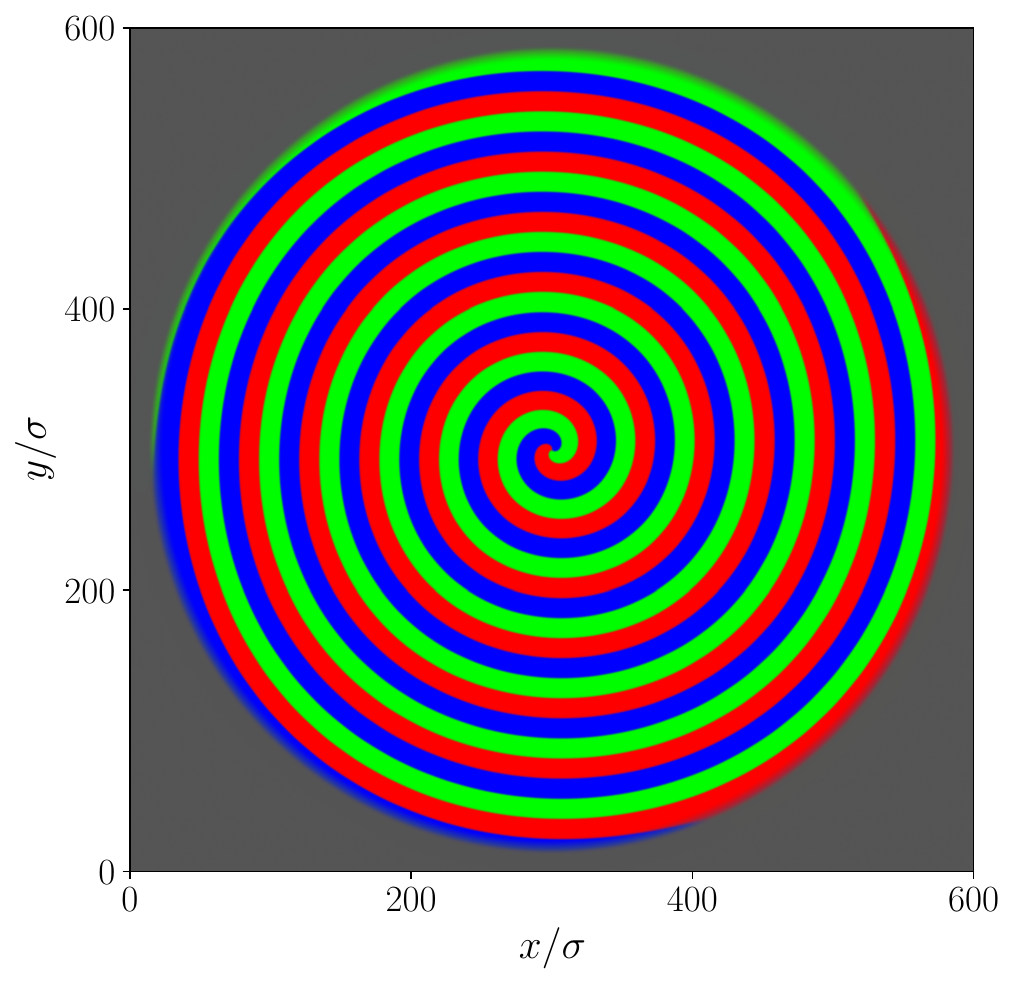}
    \caption{
        Fully developed single spiral wave obtained from direct numerical simulations of the 2D CMFEs~\eqref{eq:mf_potential}--\eqref{eq:transition_rate_contin} for $\beta z J = 7.2$ at bias parameter $\zeta = 0.5$.
        The RGB color coding is the same as in Fig.~\ref{fig:pde_snapshots}.
    }
    \label{fig:spiral_fitting}
\end{figure}

To investigate the properties of an isolated spiral wave in a finite system without boundary reflections, we apply an absorbing layer in the peripheral region that damps outgoing waves (see SM~\cite{SM}, Sec.~\ref{sm:absorbing_boundary}, for numerical details).
Under these nonreflecting conditions, the 2D CMFE simulations~\eqref{eq:mf_potential}--\eqref{eq:transition_rate_contin} sustain a steadily rotating single spiral wave as shown in Fig.~\ref{fig:spiral_fitting} for $\zeta = 0.5$ and $\beta z J = 7.2$.

To characterize the spiral geometry, we track the isophase line defined by $\arg\left[ \sum_{s=0}^2 \rho_s(\bm{x}, t) e^{2 \pi i s / 3} \right] = 0$ in the steady state, and fit the spiral arm outside the core region in polar coordinates $(r, \theta)$ to an Archimedean spiral $r(\theta) = a + b \theta$ to extract the pitch parameter $b$ and measure the steady rotational angular velocity $\Omega$.

\begin{figure}[tbp]
    \centering
    \includegraphics[width=\columnwidth]{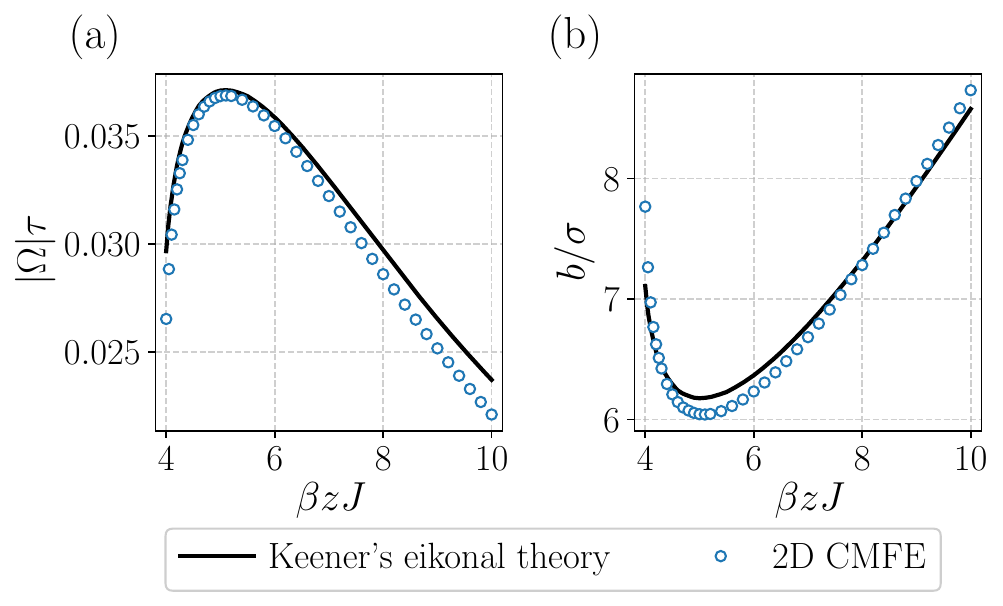}
    \caption{
        Comparison of spiral wave characteristics between theoretical predictions based on Keener's eikonal theory with the finite-core correction (solid curves) and direct 2D CMFE simulations~\eqref{eq:mf_potential}--\eqref{eq:transition_rate_contin} (blue circles) as functions of the coupling strength $\beta z J$ at bias parameter $\zeta = 0.5$.
        (a) Dimensionless rotational angular velocity $|\Omega|\tau$ and (b) dimensionless pitch parameter $b/\sigma$ of the spiral arm.
        The theoretical curves are given by Eq.~\eqref{eq:spiral_scaling_prediction} using the finite-core eigenvalue $\tilde{\Omega}^*(\tilde{r}_0)$.
    }
    \label{fig:sweep_J_dependencies}
\end{figure}

As the coupling strength $\beta z J$ is swept from $4$ to $10$ at fixed bias $\zeta = 0.5$, both the dimensionless rotational angular velocity $|\Omega|\tau$ and the dimensionless spiral pitch $b/\sigma$ exhibit nonmonotonic dependencies on $\beta z J$, as shown in Figs.~\ref{fig:sweep_J_dependencies}(a) and \ref{fig:sweep_J_dependencies}(b).

To explain these dependencies, we apply Keener's eikonal theory for spiral waves in reaction-diffusion systems~\cite{keener1986geometrical, tyson1988singular}.
When a steadily rotating spiral arm is represented in polar coordinates as $\theta(r, t) = \phi(r) - \Omega t$, the spiral geometry is governed by the radial profile $\phi(r)$.
The local tangent angle $\chi(r)$ of the spiral arm satisfies $\tan\chi = r d\phi/dr$.
Projecting the rotational velocity $r\Omega$ onto the interface normal gives $V_{\mathrm{n}} = r\Omega \cos\chi$, while the signed curvature is given by $\kappa = (d\chi/dr)\cos\chi + (\sin\chi)/r$.
Substituting these geometric relations into the eikonal equation~\eqref{eq:eikonal} and introducing dimensionless variables $\tilde{r} \coloneqq r V_{\mathrm{flat}}/D_{\mathrm{eff}}$ and $\tilde{\Omega} \coloneqq \Omega D_{\mathrm{eff}}/V_{\mathrm{flat}}^2$, we obtain the parameter-free ODE for the spiral profile~\cite{keener1986geometrical, tyson1988singular}:
\begin{equation}
    \frac{d\chi}{d\tilde{r}} = \frac{1}{\cos\chi} - \tilde{\Omega} \tilde{r} - \frac{\tan\chi}{\tilde{r}}.
    \label{eq:spiral_eikonal_ode}
\end{equation}

\begin{figure}[tbp]
    \centering
    \includegraphics[width=\columnwidth]{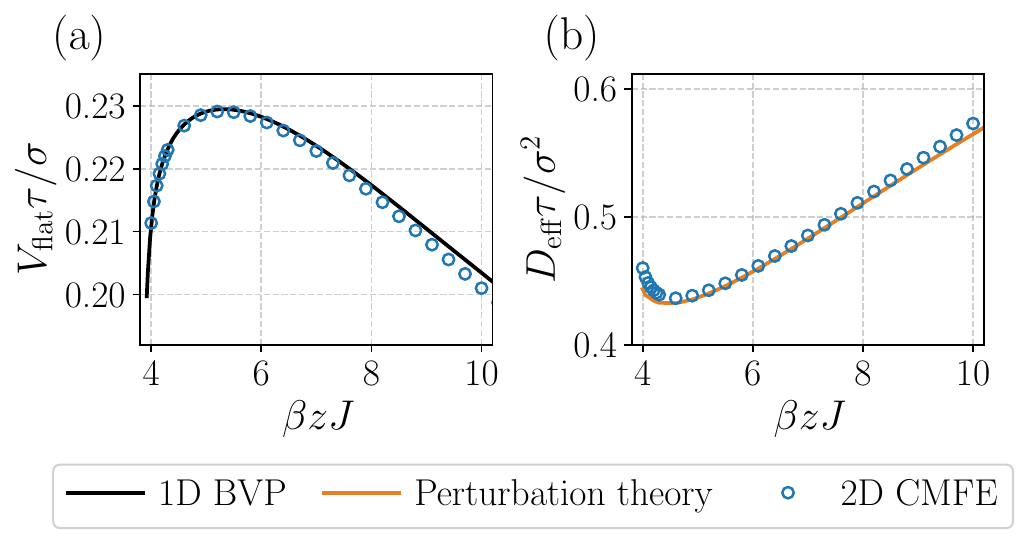}
    \caption{
        Dependence of interface dynamics on the coupling strength $\beta z J$ at bias parameter $\zeta = 0.5$.
        (a) Dimensionless flat-interface propagation velocity $V_{\mathrm{flat}}\tau/\sigma$.
        The black solid line and blue circles represent values obtained from the 1D BVP~\eqref{eq:traveling_front_ode} and direct 2D CMFE simulations~\eqref{eq:mf_potential}--\eqref{eq:transition_rate_contin}, respectively.
        (b) Dimensionless effective diffusion coefficient $D_{\mathrm{eff}}\tau/\sigma^2$.
        The orange solid line and blue circles denote theoretical predictions of the 1D perturbation theory~\eqref{eq:deff_integral_ratio} and values obtained from direct 2D CMFE simulations, respectively.
        The linear eikonal regression is restricted to small curvatures $|\kappa|\sigma \leq 0.15$.
    }
    \label{fig:circular_front_sweep_summary}
\end{figure}

In the point-core limit ($\chi(0) = 0$), this eigenvalue problem selects the universal dimensionless rotational angular velocity $\tilde{\Omega}^*(0) \simeq 0.330958$~\cite{keener1986geometrical, tyson1988singular}.
In our model, however, the spiral center forms a triple junction of domain walls.
Assuming that the core size is characterized by the finite interface width $W$ obtained from the 1D traveling front, as shown in Fig.~\ref{fig:bvp_summary}(c), we impose a finite-core boundary condition $\chi(\tilde{r}_0) = 0$ at $\tilde{r}_0 \coloneqq W V_{\mathrm{flat}} / D_{\mathrm{eff}}$, which yields a modified eigenvalue $\tilde{\Omega}^*(\tilde{r}_0)$ (see SM~\cite{SM}, Sec.~\ref{sm:finite_core_correction}).
Combining $\tilde{\Omega}^*(\tilde{r}_0)$ with $V_{\mathrm{flat}}$ and $D_{\mathrm{eff}}$ yields the theoretical predictions for the rotational angular velocity $|\Omega|$ and pitch $b$:
\begin{equation}
    |\Omega| = \tilde{\Omega}^*(\tilde{r}_0) \frac{V_{\mathrm{flat}}^2}{D_{\mathrm{eff}}}, \quad b = \frac{D_{\mathrm{eff}}}{\tilde{\Omega}^*(\tilde{r}_0) V_{\mathrm{flat}}}.
    \label{eq:spiral_scaling_prediction}
\end{equation}
As shown in Fig.~\ref{fig:sweep_J_dependencies}, the theoretical predictions of Eq.~\eqref{eq:spiral_scaling_prediction} show good quantitative agreement with the direct 2D CMFE simulations over the entire coupling range without any fitting parameters.
As shown in Fig.~\ref{fig:circular_front_sweep_summary}(b), the effective diffusion coefficient $D_{\mathrm{eff}}$ increases overall with the coupling strength $\beta z J$.
Through Eq.~\eqref{eq:spiral_scaling_prediction}, this growth of $D_{\mathrm{eff}}$, together with the decrease in $V_{\mathrm{flat}}$, accounts for the slower rotation (lower $|\Omega|$) and longer pitch (larger $b$) of the spiral waves in the strong-coupling regime.

\section{Lattice Monte Carlo Simulations}\label{sec:mc_simulations}
To compare with the CMFE simulations and theoretical analyses described in the previous sections, we perform MC simulations on triangular and square lattices ($z = 6$ and $4$) at the representative bias parameter $\zeta = 0.5$.
We employ rhombic and square simulation boxes with side length $L$ for the triangular and square lattices, respectively, containing $N = (L/\sigma)^2$ lattice sites subject to periodic boundary conditions.
The time unit $\tau_{\mathrm{MC}}$ corresponds to one Monte Carlo step, defined as $N$ attempted site updates.
In each update trial, a randomly chosen site attempts to flip its state $s$ to $s \pm 1$ with equal probability $1/2$, and the transition is accepted according to the Glauber probability~\cite{glauber1963time}.
Because each transition to $s \pm 1$ is attempted with probability $1/2$ in the MC simulation, the microscopic time unit is related to the continuum time unit by $\tau_{\mathrm{MC}} = \tau / 2$.
Statistical errors are calculated from at least three independent simulation runs.

\begin{figure}[tbp]
    \centering
    \includegraphics[width=\columnwidth]{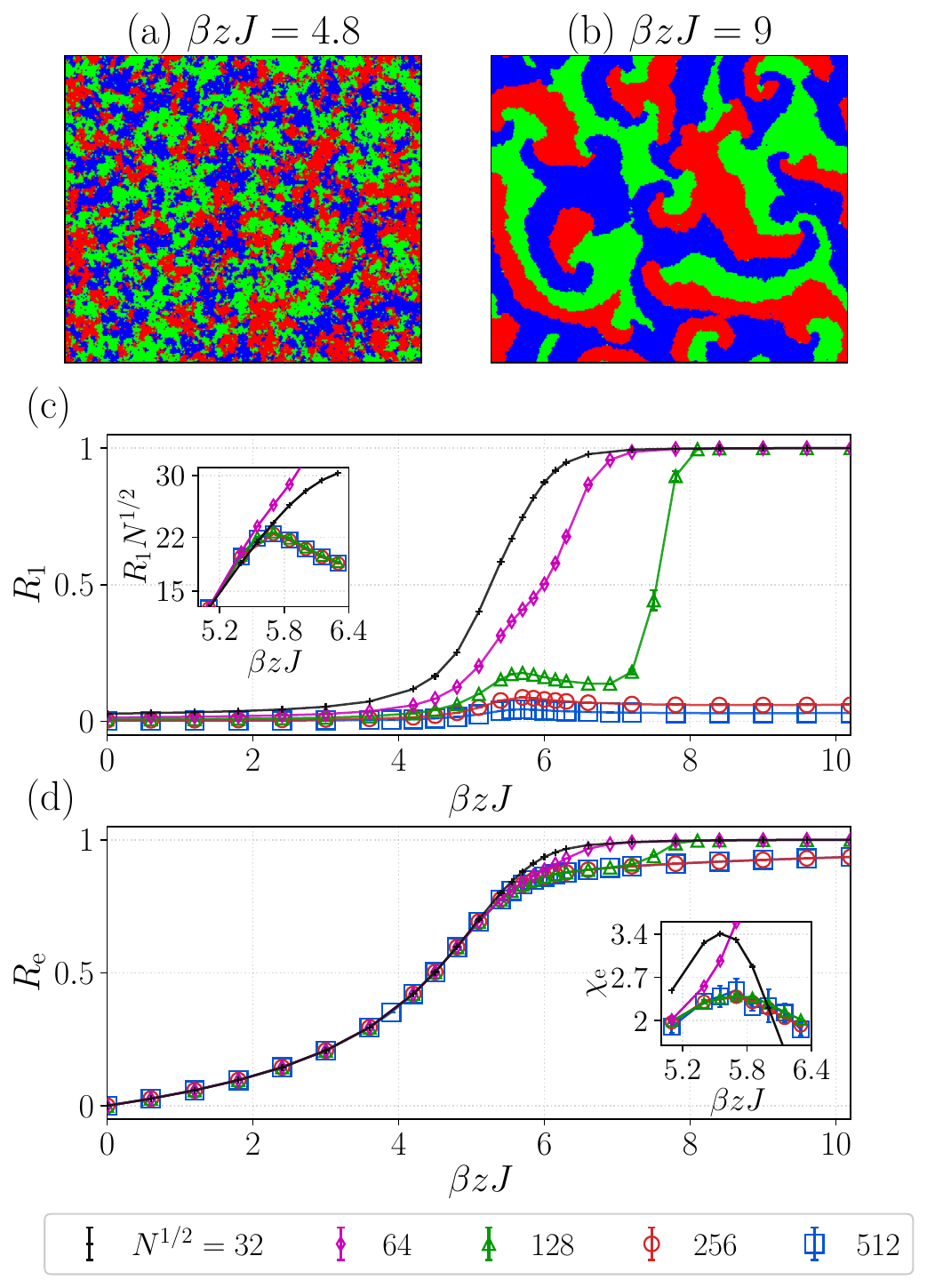}
    \caption{
        MC simulations of the three-state active Potts model on triangular lattices ($z=6$) at bias parameter $\zeta = 0.5$.
        (a), (b) Snapshots at system size $L/\sigma = 256$.
        (a) Disordered waves at $\beta z J = 4.8$.
        (b) Spiral waves at $\beta z J = 9$.
        Red, green, and blue hexagons represent the local states $s=0$, $1$, and $2$, respectively.
        (c), (d) Coupling-strength dependence of order parameters for various system sizes $L/\sigma = 32$, $64$, $128$, $256$, and $512$.
        (c) Single-state order parameter $R_1$.
        The inset shows the scaled order parameter $R_1 N^{1/2}$ near the transition.
        (d) Interfacial contact parameter $R_{\mathrm{e}}$.
        The inset shows the contact susceptibility $\chi_{\mathrm{e}}$.
        Samples trapped in the HC mode at $L/\sigma \geq 256$ are omitted from the ensemble averages for clarity.
    }
    \label{fig:mcz6}
\end{figure}

In MC simulations, disordered waves appear at small $\beta z J$, whereas spiral waves emerge at large $\beta z J$.
Snapshots of disordered and spiral waves on the triangular lattice are shown in Figs.~\ref{fig:mcz6}(a) and \ref{fig:mcz6}(b), respectively, displayed in a rectangular box of size $L \times \sqrt{3} L / 2$.
Similar wave patterns are observed on the square lattice, as shown in Fig.~\ref{fig:mcz4} in SM~\cite{SM}, Sec.~\ref{sm:mc_details}.
While the deterministic CMFEs exhibit the spatially smooth small-amplitude oscillations shown in Fig.~\ref{fig:pde_snapshots}(a), the MC simulations display the irregular domain patterns shown in Fig.~\ref{fig:mcz6}(a).
In this disordered wave regime, the contact points of the three states fluctuate and do not organize into stable spiral cores, in contrast to the fully developed spiral waves shown in Fig.~\ref{fig:mcz6}(b), corresponding to the continuum spiral waves in Fig.~\ref{fig:pde_snapshots}(c).
At even larger $\beta z J$, the homogeneous cycling (HC) mode emerges, in which the entire system is dominantly occupied by a single state that cyclically alternates via nucleation and growth as $0 \rightarrow 1 \rightarrow 2 \rightarrow 0$~\cite{noguchi2024cycling}.
While uniform solutions of the deterministic CMFEs remain permanently trapped in one of the three ordered fixed points beyond the heteroclinic bifurcation ($\beta z J > \beta z J_{\mathrm{het}}$), the HC mode in discrete lattice systems is intrinsically noise-driven, with thermal fluctuations repeatedly nucleating domains of the succeeding state $s+1$ within the dominant state $s$.
The HC mode and spiral waves alternate intermittently in smaller systems ($L/\sigma \lesssim 128$), whereas larger systems ($L/\sigma \gtrsim 256$) remain trapped in either mode over accessible simulation timescales.
Because the primary focus of this work is to elucidate interfacial dynamics and sustained spiral patterns, we restrict our subsequent analysis to the spiral wave regime.

To quantitatively characterize these spatial patterns, we introduce the single-state order parameter $R_1$ and the interfacial contact parameter $R_{\mathrm{e}}$.
The single-state order parameter $R_1$ quantifies the degree of single-state dominance~\cite{avni2025nonreciprocal, noguchi2025spatiotemporal}, corresponding to the macroscopic order parameter defined in Eq.~\eqref{eq:homogeneous_order_parameter} for the spatially homogeneous ODE system:
\begin{equation}\label{eq:r1_def}
    R_1 \coloneqq \frac{1}{N} \left\langle \left| \sum_{j=1}^N \exp\left( \frac{2\pi i s_j}{3} \right) \right| \right\rangle.
\end{equation}
The contact parameter $R_{\mathrm{e}}$ quantifies the normalized fraction of neighboring site pairs sharing the same state:
\begin{equation}
    R_{\mathrm{e}} \coloneqq -\frac{3}{zJN} \left\langle H \right\rangle - \frac{1}{2}.
\end{equation}
Here, $\langle\,\cdot\,\rangle$ denotes the time average evaluated in the steady state after discarding initial transients, combined with an ensemble average over at least three independent runs.
For an uncorrelated random configuration, both parameters vanish as $R_1 = R_{\mathrm{e}} = 0$ in the thermodynamic limit $N \to \infty$, whereas $R_1 = R_{\mathrm{e}} = 1$ for a uniformly ordered state.
In both disordered and spiral waves, the three states are almost equally distributed, yielding $R_1 \simeq 0$, whereas $R_1 \simeq 1$ in the HC mode.
In contrast, disordered and spiral waves are clearly distinguished by the contact parameter $R_{\mathrm{e}}$, which remains close to zero for disordered waves but reaches $R_{\mathrm{e}} \simeq 0.9$ for spiral waves.
For large systems ($L/\sigma \geq 128$), disordered waves smoothly transition into spiral waves as $\beta z J$ increases, whereas the HC mode coexists in smaller systems ($L/\sigma = 32$ and $64$), as shown in Figs.~\ref{fig:mcz6}(c) and \ref{fig:mcz6}(d) as well as Figs.~\ref{fig:mcz4}(c) and \ref{fig:mcz4}(d) in SM~\cite{SM}, Sec.~\ref{sm:mc_details}.
As shown in the insets of Fig.~\ref{fig:mcz6}(c) and Fig.~\ref{fig:mcz4}(c) in SM~\cite{SM}, the scaled single-state order parameter $R_1 N^{1/2}$ displays a pronounced peak at $\beta z J = 5.7$ for the triangular lattice ($z = 6$) and at $\beta z J = 6$ for the square lattice ($z = 4$).
The susceptibility $\chi_{\mathrm{e}} \coloneqq N(\langle u_{\mathrm{e}}^2 \rangle - \langle u_{\mathrm{e}} \rangle^2)$, analogous to the specific heat of the contact energy $u_{\mathrm{e}} \coloneqq -3H/(zJN) - 1/2$, also exhibits a corresponding peak at the same coupling strength, as shown in the insets of Fig.~\ref{fig:mcz6}(d) and Fig.~\ref{fig:mcz4}(d) in SM~\cite{SM}.
These coincident peaks identify this coupling threshold as the transition point between the disordered and spiral wave regimes.
Because disordered and spiral waves possess finite correlation lengths, the three states remain on average equally distributed across the system, causing the complex phase factors $\exp(2\pi i s_j / 3)$ to cancel macroscopically.
The sum in Eq.~\eqref{eq:r1_def} is therefore dominated by residual statistical fluctuations scaling as $\mathcal{O}(N^{1/2})$, which leads to $R_1 \propto N^{-1/2}$.
This finite-size decay confirms that neither dynamic pattern establishes macroscopic long-range order, in contrast to the uniformly cycling HC mode where $R_1 \sim \mathcal{O}(1)$.

\begin{figure}[tbp]
    \centering
    \includegraphics[width=\columnwidth]{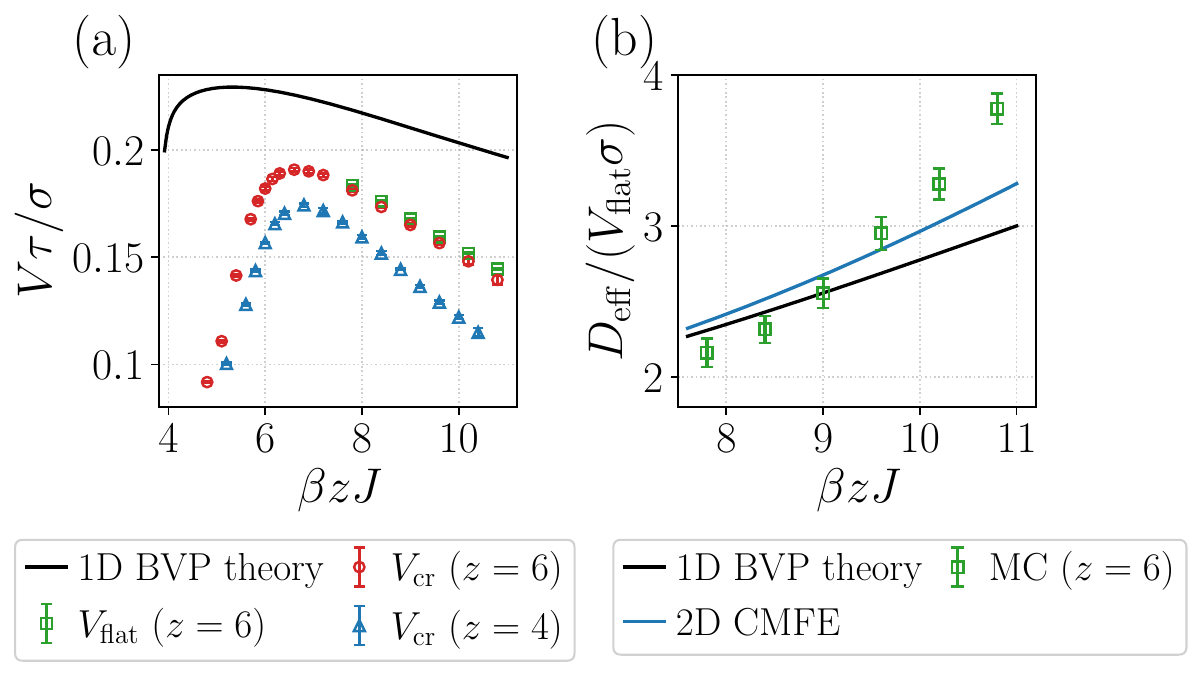}
    \caption{
        Front propagation velocity and curvature effect at bias parameter $\zeta = 0.5$.
        (a) Dimensionless front propagation velocity $V\tau/\sigma$ as a function of the coupling strength $\beta z J$.
        The black solid line represents the velocity obtained from the 1D BVP~\eqref{eq:traveling_front_ode}.
        Red open circles and blue open triangles represent the velocities $V_{\mathrm{cr}}$ extracted from spatiotemporal autocorrelations on triangular ($z=6$) and square ($z=4$) lattices, respectively.
        Green open squares represent the flat-interface velocities $V_{\mathrm{flat}}$ estimated from circular domain boundaries at $z=6$.
        (b) Dimensionless ratio $D_{\mathrm{eff}}/(V_{\mathrm{flat}}\sigma)$, corresponding to the dimensionless critical nucleus radius $R_{\mathrm{c}}/\sigma$.
        Green open squares represent the MC simulation data at $z=6$.
        The blue and black solid curves represent the results of 2D CMFE simulations and the 1D perturbation theory in Eq.~\eqref{eq:deff_integral_ratio}, respectively.
    }
    \label{fig:mcv}
\end{figure}

In MC simulations, the front propagation velocity is evaluated using two independent methods.
First, the propagation velocity of disordered and spiral waves is determined as $V_{\mathrm{cr}} \coloneqq \ell_{\mathrm{wave}}/\tau_{\mathrm{os}}$, where the wavelength $\ell_{\mathrm{wave}}$ and oscillation period $\tau_{\mathrm{os}}$ are obtained from spatial and temporal autocorrelation functions, respectively.
Second, by tracking the contraction and expansion of circular domain boundaries, the flat-interface propagation velocity $V_{\mathrm{flat}}$ and the effective diffusion coefficient $D_{\mathrm{eff}}$ in the eikonal equation~\eqref{eq:eikonal} are estimated via linear regressions, analogously to the 2D CMFE simulations.
The details of these procedures for estimating the velocity are described in SM~\cite{SM}, Sec.~\ref{sm:mc_details}, with the snapshots of circular domains and the autocorrelation functions shown in Figs.~\ref{fig:mccirc} and \ref{fig:mccrr}, respectively.
The velocities estimated by these two methods agree well with each other, as shown by comparing $V_{\mathrm{cr}}\tau/\sigma$ and $V_{\mathrm{flat}}\tau/\sigma$ in Fig.~\ref{fig:mcv}(a).
The velocity dependence on $\beta z J$ agrees well with the theoretical analysis, as seen from the data in Fig.~\ref{fig:mcv}(a).
However, the obtained peak velocity is moderately lower than that predicted by the CMFEs, which is attributed to lattice discreteness as supported by the difference observed between the triangular ($z = 6$) and square ($z = 4$) lattices.
As shown in Fig.~\ref{fig:mcv}(b), the dimensionless ratio $D_{\mathrm{eff}}/(V_{\mathrm{flat}}\sigma)$, which physically corresponds to the dimensionless critical nucleus radius $R_{\mathrm{c}}/\sigma$, shows nearly quantitative agreement with the continuum predictions, although at stronger coupling the MC simulations exhibit a steeper increase that is likewise attributed to lattice discreteness.

\section{Discussion}\label{sec:discussion}
From the perspective of continuum field theories, a distinctive feature of the CMFEs~\eqref{eq:mf_potential}--\eqref{eq:transition_rate_contin} lies in their nonlinear diffusion structure, where the spatial Laplacian appears inside the hyperbolic tangent of the Glauber transition rates~\eqref{eq:transition_rate_contin} rather than appearing as an additive linear term as in standard reaction-diffusion systems.
In the 1D traveling front from the ordered state $s = 0$ to $s = 1$, the cyclic bias drives transitions across the interface, accompanied by a localized peak of the intermediate state $s = 2$ around the interface.
However, in discrete lattice MC simulations, such an intermediate third state does not noticeably emerge within the domain boundary, indicating that the third component itself plays only a minor role in determining the propagation velocity.
Instead, front propagation in both continuum and discrete systems is fundamentally driven by cyclic state transitions along the loop $s = 0 \rightarrow 1 \rightarrow 2 \rightarrow 0$ across the interface.
The localized peak of the intermediate third state is closely related to the wave dynamics observed in six-state active Potts models~\cite{noguchi2026spiral}, where spiral waves dominated by three alternating states (e.g., $s = 0$, $2$, and $4$) advance through two-step transitions $s \rightarrow s+1 \rightarrow s+2$ via intermediate states nucleated at domain boundaries.
Remarkably, despite the nonlinear diffusion coupling and the localized intermediate-state peak around the interface, the 2D curved domain-wall dynamics in this model is universally reduced to the standard geometric eikonal equation~\eqref{eq:eikonal}, $V_{\mathrm{n}} = V_{\mathrm{flat}} - D_{\mathrm{eff}} \kappa$, which successfully captures the essential driving mechanism bridging microscopic cyclic state transitions with macroscopic interface motion.
Beyond steady wave rotation, this geometric framework also provides a natural criterion for the critical nucleus radius $R_{\mathrm{c}} = D_{\mathrm{eff}}/V_{\mathrm{flat}}$, elucidating how curvature-dependent front propagation dictates domain growth and collapse.

Regarding the regime of wave pattern formation, while the CGL equation derived in Sec.~\ref{sec:cgl} describes small-amplitude dynamics near the Hopf bifurcation as in cyclic predator-prey models~\cite{szczesny2013does, szczesny2014characterization}, the three-state active Potts model sustains stable spiral waves predominantly in the large-amplitude regime extending far beyond this narrow validity window.
When the coupling strength $\beta z J$ exceeds the heteroclinic bifurcation point $\beta z J_{\mathrm{het}}$, the limit-cycle orbit collapses via collision with saddle points in the homogeneous phase space, extinguishing uniform temporal oscillations.
In this strong-coupling regime, large-amplitude spiral waves are sustained not by weakly nonlinear CGL dynamics, but by $\mathbb{Z}_3$-symmetric propagating domain walls connecting the three coexisting bulk ordered states around a central topological defect.
Consequently, the macroscopic spiral geometry and rotation are governed not by amplitude equations, but by the geometric eikonal framework formulated in this work.
This framework enables the rotational angular velocity and pitch of spiral waves to be predicted solely from the flat propagation velocity $V_{\mathrm{flat}}$ and the effective diffusion coefficient $D_{\mathrm{eff}}$ of the 1D traveling front.
In particular, whereas the CGL equation near the Hopf onset yields only monotonic scaling with the coupling strength, the nonmonotonic dependencies of the front velocity and the resulting spiral wave characteristics on $\beta z J$ arise inherently from this large-amplitude domain-wall dynamics.

Next, we discuss the validity and inherent limitations of the spatial continuum mean-field approximation employed in this study.
Generally, mean-field theories for discrete lattice models neglect local fluctuations and spatial correlations, which tends to overestimate the stability of ordered phases, shift the critical transition points, and fail to capture non-mean-field critical exponents, especially in two dimensions~\cite{goldenfeld2018lectures}.
As observed in our comparison with MC simulations in Sec.~\ref{sec:mc_simulations}, the continuum theory accurately predicts the deterministic front propagation velocity and curvature-dependent dynamics in the moderate-coupling regime.
At small $\beta z J$, however, stochastic fluctuations dominate in the MC simulations, leading to disordered waves instead of the smooth small-amplitude waves predicted by the CMFEs.
Conversely, in the strong-coupling regime, the effects of lattice discreteness become prominent, resulting in lower front propagation velocities than the continuum predictions.
Developing continuum field-theoretic formulations that incorporate stochastic noise terms or pair-correlation approximations beyond local mean-field theories~\cite{blom2023thermodynamically, blom2025local} would be an important subject for future studies to quantitatively account for fluctuation-induced phenomena.

The continuum framework and interface theory developed in this work have broad implications for nonequilibrium systems.
Three-state cyclic driven dynamics serves as a minimal continuum model for rock-paper-scissors games and systems with cyclic dominance~\cite{kerr2002local, reichenbach2007mobility, frey2010evolutionary, szczesny2013does, szczesny2014characterization, szolnoki2014cyclic, dobramysl2018stochastic, griffin2024spatial}.
In related contexts, coupled cyclic transition networks exhibit non-Hermitian topological phenomena~\cite{knebel2020topological, yoshida2022non, wang2026robust}.
Furthermore, traveling pattern formation driven by such cyclic state transitions shares fundamental dynamical features with pattern formation in nonreciprocal continuum models, such as Allen--Cahn~\cite{liu2023non}, Cahn--Hilliard~\cite{saha2020scalar, you2020nonreciprocity, frohoff2021suppression, frohoff2021localized, frohoff2023non, frohoff2023nonreciprocal, suchanek2023entropy, greve2024amplitude, saha2024phase, brauns2024nonreciprocal, greve2025coexistence}, and Swift--Hohenberg equations~\cite{fruchart2021non, tateyama2024pattern, tateyama2026higher}.
While two-component nonreciprocal Ising models can be mapped onto four-state active clock models via an Ashkin--Teller-type transformation~\cite{avni2025nonreciprocal, avni2025dynamical, noguchi2025dynamic}, active Potts models with three or more states represent an intrinsically generalized class of internally driven media where spatial interactions remain reciprocal.
Uncovering potential mathematical correspondences between such multistate active systems and multicomponent nonreciprocal field theories offers an intriguing avenue for future theoretical research.
In addition to well-established frameworks for self-propelled particles and continuum active fluids~\cite{ramaswamy2010mechanics, marchetti2013hydrodynamics, ramaswamy2017active}, investigating systems driven by internal state transitions provides a valuable perspective for nonequilibrium physics.
The multistate lattice formulation and geometric interface theory developed in this work provide a versatile framework for exploring domain dynamics and wave propagation in such internally driven nonequilibrium media.

\section{Conclusion}\label{sec:conclusion}
In conclusion, we have presented a continuum field theory for the three-state active Potts model with a cyclic transition bias.
By deriving the CMFEs from the microscopic master equation, we identified the nonequilibrium dynamical regimes spanning from weakly nonlinear oscillations near the Hopf bifurcation to large-amplitude spiral waves sustained by $\mathbb{Z}_3$-symmetric propagating domain walls connecting the three coexisting ordered states.

Furthermore, we formulated an eikonal interface theory that consistently bridges the microscopic parameters and the macroscopic interface motion.
We found that the propagation velocity of flat fronts exhibits a maximum at a relatively low $\beta z J$ in the large-amplitude spiral wave regime.
Combining the 1D traveling front solution with the theoretical effective diffusion coefficient, we semianalytically predicted the steady rotational angular velocity and pitch of 2D spiral waves.
These theoretical predictions for the spiral geometry were quantitatively confirmed by 2D CMFE simulations, while the predictions for front velocity and curvature response were validated across both continuum models and discrete lattice MC simulations.

The framework developed in this work provides a unified perspective bridging microscopic cyclic state transitions, weakly nonlinear oscillations, and geometric interface dynamics.
Our approach offers a basis for exploring nonreciprocal dynamics, biological pattern selection in cyclic games, and pattern formation in multicomponent nonequilibrium media.

\section*{Acknowledgment}
We thank Kristian Blom, Uwe Thiele, and Shigeyuki Komura for fruitful discussions.
This work was supported by the JSPS Core-to-Core Program ``Advanced core-to-core network for the physics of self-organizing active matter'' (JPJSCCA20230002) and MEXT Promotion of Distinctive Joint Usage/Research Center Support Program Grant Number JPMXP0724020292.
The simulations were partially carried out at ISSP Supercomputer Center, University of Tokyo (ISSPkyodo-SC-2026-0075).
Y.T.\ acknowledges support from the Japan Science and Technology Agency (JST) SPRING (No.~JPMJSP2109) and the Cooperative Research Program of ``Network Joint Research Center for Materials and Devices'' (No.~20265001).
H.K.\ acknowledges support from JSPS KAKENHI Grant Numbers JP25K00918 and JP26K00671, MEXT KAKENHI Grant Number JP26H00384, and the Cooperative Research Program of ``Network Joint Research Center for Materials and Devices'' (No.~20264003).
H.N.\ acknowledges support from JSPS KAKENHI Grant Number JP24K06973.

\appendix
\section{Derivation of the Linearized Equation and Nonlinear Coefficients}\label{app:derivation_equations}
\subsection{Derivation of the Linear Dispersion Relation}\label{app:linear_dispersion}
This appendix provides the detailed mathematical derivation of the linearized evolution equation~\eqref{eq:master_eq_contin_linearized} and the linear dispersion relation~\eqref{eq:dispersion_relation} presented in Sec.~\ref{sec:linear_dispersion}.
We substitute the spatiotemporal perturbation $\rho_s(\bm{x}, t) = 1/3 + \delta\rho_s(\bm{x}, t)$ into the CMFEs~\eqref{eq:mf_potential}--\eqref{eq:transition_rate_contin}.
First, we linearize the continuous transition rates $w(s \rightarrow s \pm 1; \bm{x})$ in Eq.~\eqref{eq:transition_rate_contin}.
We express the potential difference as $U_{s\pm 1} - U_s = \delta U_{s\pm 1} - \delta U_s$, where $\delta U_s$ is the potential perturbation from the homogeneous disordered state.
Expanding the hyperbolic tangent function in Eq.~\eqref{eq:transition_rate_contin} to first order with respect to the potential differences, we obtain
\begin{equation}
    w(s \rightarrow s \pm 1) \simeq \frac{1}{2\tau} \left[ 1 \pm \zeta - \frac{\beta(1-\zeta^2)}{2} \left( \delta U_{s\pm 1} - \delta U_s \right) \right],
    \label{eq:transition_rate_linearized_app}
\end{equation}
where $\zeta = \tanh(\beta h / 2)$ and we have used $\sech^2(\beta h / 2) = 1 - \zeta^2$.

Next, we substitute the perturbation and the linearized rates~\eqref{eq:transition_rate_linearized_app} into the master equation~\eqref{eq:master_eq_contin}.
The inflow and outflow terms are evaluated to the first order as
\begin{align}
    w(s \mp 1 \rightarrow s) \rho_{s \mp 1} &\simeq \frac{1}{2\tau} \biggl[ \frac{1 \pm \zeta}{3} + (1 \pm \zeta)\delta\rho_{s \mp 1} \notag\\
    &\quad - \frac{\beta(1-\zeta^2)}{6} \left( \delta U_s - \delta U_{s \mp 1} \right) \biggr],
\end{align}
and
\begin{align}
    &\left[ w(s \rightarrow s+1) + w(s \rightarrow s-1) \right] \rho_s \notag\\
    &\simeq \frac{1}{2\tau} \biggl[ \frac{2}{3} + 2\delta\rho_s - \frac{\beta(1-\zeta^2)}{6} \left( \delta U_{s+1} + \delta U_{s-1} - 2\delta U_s \right) \biggr].
\end{align}
Substituting these expressions into Eq.~\eqref{eq:master_eq_contin}, we obtain the linearized evolution equation~\eqref{eq:master_eq_contin_linearized}.

To derive the linear dispersion relation, we substitute the normal-mode ansatz $\delta\rho_s(\bm{x}, t) = \delta\tilde{\rho} e^{\lambda t} e^{i(\bm{q}\cdot\bm{x} + 2\pi s / 3)} + \mathrm{c.c.}$ into Eq.~\eqref{eq:master_eq_contin_linearized}, which yields the eigenvalue equation for the complex growth rate $\lambda(q)$:
\begin{align}
    \tau \lambda(q) &= -1 + \frac{1+\zeta}{2} e^{-2\pi i / 3} + \frac{1-\zeta}{2} e^{2\pi i / 3} \notag\\
    &\quad + \frac{(1-\zeta^2)\beta z J}{6} \left( 1 - \frac{\sigma^2 q^2}{4} \right) \notag\\
    &\qquad \times \left( 2 - e^{-2\pi i / 3} - e^{2\pi i / 3} \right).
    \label{eq:app_dispersion_algebraic}
\end{align}
By Euler's formula $e^{\pm 2\pi i / 3} = -1/2 \pm i\sqrt{3}/2$, the transition rate terms evaluate to
\begin{equation}
    -1 + \frac{1+\zeta}{2} e^{-2\pi i / 3} + \frac{1-\zeta}{2} e^{2\pi i / 3} = -\frac{3}{2} - i \frac{\sqrt{3}}{2} \zeta,
\end{equation}
and the potential coupling term simplifies to
\begin{equation}
    2 - e^{-2\pi i / 3} - e^{2\pi i / 3} = 3.
\end{equation}
Substituting these relations, we obtain Eq.~\eqref{eq:dispersion_relation}.

\subsection{Derivation of the Coefficients of the Complex Amplitude Equation}\label{app:amplitude_coefficients}
This appendix details the derivation of the nonlinear expansion coefficients for the complex amplitude equation described in Sec.~\ref{sec:cgl}.
To derive the lowest-order nonlinear coefficients $u_2$ and $u_3$, we expand the continuous transition rates $w(s \rightarrow s \pm 1; \bm{x})$ in Eq.~\eqref{eq:transition_rate_contin} up to the third order with respect to the potential differences.
For a spatially homogeneous system near the Hopf bifurcation point, $J \simeq J_{\mathrm{Hopf}}$, the potential perturbation is given by $\delta U_s = -z J_{\mathrm{Hopf}} \delta\rho_s$.
We expand the transition rates as
\begin{align}
    w(s \rightarrow s \pm 1) &\simeq \frac{1}{2\tau} \biggl[ 1 \pm \zeta - \frac{\beta(1-\zeta^2)}{2} \delta U_{s, s \pm 1} \notag\\
    &\phantom{\simeq \frac{1}{2\tau} \biggl[\ } \mp \frac{\beta^2 \zeta(1-\zeta^2)}{4} \delta U_{s, s \pm 1}^2 \notag\\
    &\phantom{\simeq \frac{1}{2\tau} \biggl[\ } - \frac{\beta^3(1-\zeta^2)(1-3\zeta^2)}{24} \delta U_{s, s \pm 1}^3 \biggr],
\end{align}
where $\delta U_{s, s'} \coloneqq \delta U_{s'} - \delta U_s = -z J_{\mathrm{Hopf}} (\delta\rho_{s'} - \delta\rho_s)$.
Substituting the amplitude expansion $\delta\rho_s = A e^{2\pi i s / 3} + \mathrm{c.c.}$ and using $\beta z J_{\mathrm{Hopf}} = 3 / (1 - \zeta^2)$, we evaluate the potential differences as
\begin{equation}
    \beta \delta U_{s, s \pm 1} = -\frac{3}{1-\zeta^2} \left[ A e^{2\pi i s / 3} ( e^{\pm 2\pi i / 3} - 1 ) + \mathrm{c.c.} \right].
\end{equation}
Substituting these rates into the master equation~\eqref{eq:master_eq_contin} and collecting terms up to the third order of $A$ and $A^*$, we express the time derivative $\tau \partial_t \delta\rho_s$ as a power series of the amplitude.
To project this equation onto the unstable mode, we multiply both sides by $e^{-2\pi i s / 3}$ and sum over the three states $s = 0$, $1$, and $2$, using the discrete orthogonality relation $\sum_{s=0}^2 e^{2\pi i s (m-n) / 3} = 3 \delta_{m, n}$.
Evaluating the linear, quadratic, and cubic terms of the projection, we obtain the evolution equation for the complex amplitude:
\begin{equation}
    \tau \partial_t A = u_1 A + u_2 (A^*)^2 + u_3 |A|^2 A + \mathcal{O}(|A|^4),
\end{equation}
where the linear coefficient is $u_1 = \mu + i\tau\omega_0$.
The quadratic and cubic coefficients are computed as
\begin{equation}
    u_2 = \frac{9}{4} + i \frac{9 \sqrt{3} \zeta}{4 (1 - \zeta^2)}, \quad
    u_3 = \frac{81(3\zeta^2-1)}{8(1-\zeta^2)^2} + i \frac{27\sqrt{3}\zeta}{8(1-\zeta^2)}.
\end{equation}
Through the near-identity transformation detailed in Appendix~\ref{app:5th_order_nit}, these coefficients yield the effective Landau coefficient $g$ in the CGL equation~\eqref{eq:cgl_equation}.

\section{Fifth-Order Amplitude Equation and Complex Ginzburg--Landau Reduction via Near-Identity Transformation}\label{app:5th_order_nit}
This appendix provides the mathematical formulation of the fifth-order weakly nonlinear analysis and the near-identity transformation (NIT) supporting the CGL reduction in Sec.~\ref{sec:cgl}.
To determine whether the primary Hopf bifurcation is supercritical or subcritical and evaluate the amplitude saturation in the vicinity of the Bautin point $\zeta_{\mathrm{c}} = 1/\sqrt{3}$, we extend the perturbative expansion up to the fifth order and perform the near-identity transformation.

\subsection{Fifth-Order Amplitude Equation and Projection Coefficients}\label{app:5th_order_expansion}
Following the third-order expansion procedure described in Appendix~\ref{app:amplitude_coefficients}, we expand the continuous transition rates $w(s \rightarrow s \pm 1; \bm{x})$ in Eq.~\eqref{eq:transition_rate_contin} up to the fifth order with respect to the potential differences $\delta U_{s, s \pm 1}$.
Substituting these rates into the master equation~\eqref{eq:master_eq_contin} and collecting terms up to the fifth order in the complex amplitude $A$ and its complex conjugate $A^*$, the time evolution of the critical amplitude obeys
\begin{align}
    \tau \partial_t A &= \gamma \nabla^2 A + u_1 A + u_2 (A^*)^2 + u_3 |A|^2 A \notag\\
    &\quad + u_{4,(4,0)} A^4 + u_{4,(1,3)} A (A^*)^3 \notag\\
    &\quad + u_{5,(3,2)} |A|^4 A + u_{5,(0,5)} (A^*)^5 + \mathcal{O}(|A|^6),
    \label{eq:5th_order_A_ode_app}
\end{align}
where each coefficient $u_{d,(p,q)}$ of the monomial $A^p (A^*)^q$ with degree $d = p + q$ is obtained by projecting the expanded CMFEs onto the primary critical mode:
\begin{subequations}
\label{eq:u_coefficients}
\begin{align}
    &u_1 = \mu - i \frac{\sqrt{3} \zeta}{2},
    \\
    &u_2 = \frac{3 \left(2 \mu + 3\right)}{4} + i \frac{3 \sqrt{3} \zeta \left(4 \mu + 3\right)}{4 \left(1 - \zeta^{2}\right)},
    \\
    &u_3 = \frac{81 \left(2 \mu + 1\right) \left(3 \zeta^{2} - 1\right)}{8 \left(1 - \zeta^{2}\right)^{2}} + i \frac{9 \sqrt{3} \zeta \left(4 \mu + 3\right)}{8 \left(1 - \zeta^{2}\right)},
    \\
    &u_{4,(4,0)} = \frac{81 \left(3 \zeta^{2} - 1\right)}{16 \left(1 - \zeta^{2}\right)^{2}} - i \frac{81 \sqrt{3} \zeta \left(3 \zeta^{2} - 2\right)}{16 \left(1 - \zeta^{2}\right)^{3}},
    \\
    &u_{4,(1,3)} = \frac{81 \left(3 \zeta^{2} - 1\right)}{8 \left(1 - \zeta^{2}\right)^{2}} + i \frac{81 \sqrt{3} \zeta \left(3 \zeta^{2} - 2\right)}{4 \left(1 - \zeta^{2}\right)^{3}},
    \\
    &u_{5,(3,2)} = \frac{729 \left(15 \zeta^{4} - 15 \zeta^{2} + 2\right)}{16 \left(1 - \zeta^{2}\right)^{4}} + i \frac{243 \sqrt{3} \zeta \left(3 \zeta^{2} - 2\right)}{16 \left(1 - \zeta^{2}\right)^{3}},
    \\
    &u_{5,(0,5)} = - \frac{729 \left(15 \zeta^{4} - 15 \zeta^{2} + 2\right)}{160 \left(1 - \zeta^{2}\right)^{4}} + i \frac{243 \sqrt{3} \zeta \left(3 \zeta^{2} - 2\right)}{32 \left(1 - \zeta^{2}\right)^{3}}.
\end{align}
\end{subequations}
Under the cyclic state permutation $s \rightarrow s + 1$, the complex amplitude transforms as $A \rightarrow A e^{2\pi i / 3}$.
Consequently, the discrete $\mathbb{Z}_3$ equivariance permits all monomials $A^p (A^*)^q$ in Eq.~\eqref{eq:5th_order_A_ode_app} satisfying $p - q - 1 \equiv 0 \pmod 3$.
However, in addition to the resonant terms $|A|^2 A$ and $|A|^4 A$ satisfying the exact $\mathrm{U}(1)$ phase-translation symmetry condition $p - q - 1 = 0$, Eq.~\eqref{eq:5th_order_A_ode_app} also contains nonresonant terms [$(A^*)^2$, $A^4$, $A(A^*)^3$, and $(A^*)^5$], which break continuous $\mathrm{U}(1)$ symmetry.

\subsection{Near-Identity Transformation and Normal Form Complex Ginzburg--Landau Equation}\label{app:nit_reduction}
To eliminate these $\mathbb{Z}_3$-equivariant yet nonresonant terms and reduce the amplitude equation to the proper CGL equation with continuous $\mathrm{U}(1)$ phase-translation symmetry, we introduce the following fifth-order NIT:
\begin{align}
    A &= B + c_{2,(0,2)} (B^*)^2 + c_{4,(1,3)} B (B^*)^3 \notag\\
    &\quad + c_{4,(4,0)} B^4 + c_{5,(0,5)} (B^*)^5 + \mathcal{O}(|B|^6).
\end{align}
The near-identity transformation contains only the nonresonant monomials, while the resonant terms satisfying $p - q - 1 = 0$ cannot be removed and are retained in the normal form equation.
Each transformation coefficient $c_{d,(p,q)}$ is determined recursively at each order by solving the decoupled homological equations~\cite{kuznetsov2023elements,wiggins2003introduction}.
To demonstrate the mathematical steps of this recursive process, we express the NIT in the general series form:
\begin{equation}
    A = B + \sum_{d \geq 2} \sum_{p+q=d} c_{d,(p,q)} B^p (B^*)^q.
\end{equation}
We substitute this transformation into the amplitude equation~\eqref{eq:5th_order_A_ode_app} and postulate the normal form equation for the transformed amplitude $B$:
\begin{equation}
    \tau \partial_t B = u_1 B + \sum_{d \geq 2} \sum_{p+q=d} g_{d,(p,q)} B^p (B^*)^q,
\end{equation}
where the summation runs over the resonant terms ($p - q - 1 = 0$) and the coefficients depend on the bifurcation parameter $\mu$.
Matching the terms of order $d$ on both sides of the transformed equation, we obtain the algebraic relation for each $(p,q)$ term at order $d = p + q$:
\begin{equation}
    i \tau \omega_0 (p - q - 1) c_{d,(p,q)} + g_{d,(p,q)} = R_{d,(p,q)},
    \label{eq:homological_equation_app}
\end{equation}
where $R_{d,(p,q)}$ represents the coefficient of the $B^p (B^*)^q$ term at order $d$ generated by the substitution of the lower-order transformation terms, with $R_{2,(p,q)} = u_{2,(p,q)}$ at order $d=2$.
Because the right-hand side $R_{d,(p,q)}$ depends only on the original coefficients $u_{d,(p,q)}$ and the lower-order transformation coefficients $c_{k,(p',q')}$ with $k < d$ and $p'+q'=k$, it is completely determined recursively.

From the homological equation~\eqref{eq:homological_equation_app}, the coefficients are determined depending on whether the term is resonant.
If the term is resonant, namely, $p - q - 1 = 0$, which corresponds to $(p, q) = (2, 1)$ at $d=3$, and $(p, q) = (3, 2)$ at $d=5$, the first term in Eq.~\eqref{eq:homological_equation_app} vanishes, determining the normal form coefficients:
\begin{equation}
    g_{d,(q+1,q)} = R_{d,(q+1,q)}.
\end{equation}
Otherwise, for the nonresonant terms where $p - q - 1 \neq 0$, we set $g_{d,(p,q)} = 0$ to eliminate them, yielding the transformation coefficients:
\begin{equation}
    c_{d,(p,q)} = \frac{R_{d,(p,q)}}{i \tau \omega_0 (p - q - 1)}.
\end{equation}
Solving these homological equations recursively, we obtain the transformation coefficients as follows:
\begin{subequations}
\label{eq:nit_coefficients}
\begin{align}
    c_{2,(0,2)} &= \frac{4 \mu + 3}{2 \left(1 - \zeta^{2}\right)} - i \frac{\sqrt{3} \left(2 \mu + 3\right)}{6 \zeta}, \\
    c_{4,(1,3)} &= \frac{9 \left(7 \zeta^{4} - 2 \zeta^{2} + 1\right)}{2 \left(1 - \zeta^{2}\right)^{3} \zeta^{2}} - i \frac{3 \sqrt{3} \left(5 \zeta^{4} + 5 \zeta^{2} + 2\right)}{4 \left(1 - \zeta^{2}\right)^{2} \zeta^{3}}, \\
    c_{4,(4,0)} &= - \frac{9 \left(10 \zeta^{2} - 1\right)}{4 \left(1 - \zeta^{2}\right)^{2} \zeta^{2}} \notag \\
    &\quad - i \frac{3 \sqrt{3} \left(11 \zeta^{6} - 45 \zeta^{4} + 15 \zeta^{2} + 1\right)}{8 \left(1 - \zeta^{2}\right)^{3} \zeta^{3}}, \\
    c_{5,(0,5)} &= - \frac{9 \left(64 \zeta^{8} - 463 \zeta^{6} + 426 \zeta^{4} - 61 \zeta^{2} - 2\right)}{32 \left(1 - \zeta^{2}\right)^{4} \zeta^{4}} \notag \\
    &\quad + i \frac{9 \sqrt{3} \left(730 \zeta^{6} - 1605 \zeta^{4} + 648 \zeta^{2} - 25\right)}{160 \left(1 - \zeta^{2}\right)^{4} \zeta^{3}}.
\end{align}
\end{subequations}
Applying this transformation and incorporating all nonresonant terms into higher-order contributions, we obtain the fifth-order proper CGL equation for the complex amplitude $B$:
\begin{align}
    \tau \partial_t B &= \gamma \nabla^2 B + (\mu + i\tau\omega_0) B \notag\\
    &\quad + g_3 |B|^2 B + g_5 |B|^4 B + \mathcal{O}(|B|^6),
\end{align}
where the Landau coefficients $g_3$ and $g_5$ are given by
\begin{subequations}
\label{eq:landau_coefficients_5th}
\begin{align}
    g_3 &= \frac{81 \left(2 \mu + 1\right) \left(3 \zeta^{2} - 1\right)}{8 \left(1 - \zeta^{2}\right)^{2}} \notag \\
    &\quad - i \frac{3 \sqrt{3} \left(4 \mu \zeta^{4} - 44 \mu \zeta^{2} - 8 \mu + 3 \zeta^{4} - 15 \zeta^{2} - 6\right)}{8 \left(1 - \zeta^{2}\right)^{2} \zeta}, \\
    g_5 &= \frac{243 \left(51 \zeta^{6} - 51 \zeta^{4} + 13 \zeta^{2} - 1\right)}{16 \left(1 - \zeta^{2}\right)^{4} \zeta^{2}} \notag \\
    &\quad - i \frac{27 \sqrt{3} \left(11 \zeta^{8} - 206 \zeta^{6} + 105 \zeta^{4} - 14 \zeta^{2} - 4\right)}{16 \left(1 - \zeta^{2}\right)^{4} \zeta^{3}}.
\end{align}
\end{subequations}
Note that at the Bautin point $\zeta_{\mathrm{c}} = 1/\sqrt{3} \simeq 0.577$ ($\beta h_{\mathrm{c}} = \ln(2 + \sqrt{3}) \simeq 1.32$), the real part of the fifth-order coefficient is $\mathrm{Re}(g_5) = -6561/64 < 0$, which serves as a stabilizing contribution that saturates the amplitude.
In the immediate vicinity above the Bautin point ($\zeta \gtrsim \zeta_{\mathrm{c}}$), where $\mathrm{Re}(g_3) > 0$ and $\mathrm{Re}(g_5) < 0$, the unstable limit cycle emerging from the subcritical Hopf bifurcation can undergo a saddle-node bifurcation of limit cycles to form a stable limit cycle before colliding with the saddles.
As $\zeta$ increases further, however, the unstable limit cycle in the spatially homogeneous system collides directly with the $\mathbb{Z}_3$-symmetric saddles and disappears via the heteroclinic bifurcation, well before reaching the higher-codimension point with $\mathrm{Re}(g_5) = 0$ at $\zeta \simeq 0.811$ ($\beta h \simeq 2.26$).
In any case, because the macroscopic rotating spiral waves in two dimensions are sustained by propagating domain walls connecting the coexisting bulk ordered states rather than by uniform oscillations, these homogeneous limit-cycle bifurcations do not affect the spiral wave dynamics.

\section{Perturbation Theory for Curved Interfaces and Adjoint Operator}\label{app:curvature_perturbation_theory}
This appendix details the perturbation theory for curved interfaces and the adjoint solvability analysis utilized in Sec.~\ref{sec:circular_front}.
We present the algebraic details of the perturbation expansion for the 2D CMFEs~\eqref{eq:mf_potential}--\eqref{eq:transition_rate_contin} and derive the solvability condition for the effective diffusion coefficient $D_{\mathrm{eff}}$.

\subsection{Solvability Condition for Effective Diffusion Coefficient}\label{app:curvature_solvability}
Due to the conservation of probability $\sum_{s=0}^2 \rho_s(\bm{x}, t) = 1$, the density field is fully described by the two-component vector $\bm{\rho} \coloneqq (\rho_0, \rho_1)^\top$, with the third component eliminated via $\rho_2 = 1 - \rho_0 - \rho_1$.
In comoving polar coordinates $(\xi, \theta)$ attached to an axisymmetric circular interface with local curvature $\kappa = 1/\xi$, the spatial Laplacian is expanded as $\nabla^2 = \partial_\xi^2 + \kappa \partial_\xi$.
We expand the traveling front profile and normal propagation velocity in powers of the local curvature as $\bm{\rho}(\xi) = \bm{\rho}^{(0)}(\xi) + \kappa \bm{\rho}^{(1)}(\xi) + \mathcal{O}(\kappa^2)$ and $V_{\mathrm{n}} = V_{\mathrm{flat}} - D_{\mathrm{eff}} \kappa + \mathcal{O}(\kappa^2)$, where the unperturbed 1D traveling front solution $\bm{\rho}^{(0)}(\xi)$ satisfies the ODE system~\eqref{eq:traveling_front_ode} with steady propagation velocity $V_{\mathrm{flat}}$.
Substituting these expansions into the CMFEs~\eqref{eq:mf_potential}--\eqref{eq:transition_rate_contin} and collecting terms of order $\mathcal{O}(\kappa)$, we obtain the linear inhomogeneous differential equation for the perturbation vector $\bm{\rho}^{(1)}$:
\begin{equation}
    \mathcal{J} \bm{\rho}^{(1)} = - D_{\mathrm{eff}} \partial_\xi \bm{\rho}^{(0)} + \bm{N}(\bm{\rho}^{(0)}),
    \label{eq:app_linear_inhomogeneous_ode}
\end{equation}
where $\mathcal{J} \coloneqq -V_{\mathrm{flat}} \partial_\xi + \mathsf{P}(\xi) + \mathsf{Q}(\xi)\partial_\xi^2$ is the linearized Jacobian operator.
Here, the $2 \times 2$ coefficient matrices $\mathsf{P}(\xi)$ and $\mathsf{Q}(\xi)$ are obtained by expanding the transition terms around the unperturbed 1D front $\bm{\rho}^{(0)}(\xi)$ with respect to $\bm{\rho}^{(1)}$ and $\partial_\xi^2 \bm{\rho}^{(1)}$, respectively.
Their explicit expressions are given below in Appendix~\ref{app:curvature_algebraic_details}.

Due to translational symmetry, $\mathcal{J}$ possesses a zero eigenvalue corresponding to the Goldstone mode $\partial_\xi \bm{\rho}^{(0)}(\xi)$ satisfying $\mathcal{J} \partial_\xi \bm{\rho}^{(0)} = 0$.
Because $\mathcal{J}$ is non-self-adjoint under the standard $L^2$ inner product $\langle \bm{u}, \bm{v} \rangle \coloneqq \int_{-\infty}^\infty \bm{u}(\xi) \cdot \bm{v}(\xi) d\xi$ with $\bm{u} \cdot \bm{v} \coloneqq \sum_{s=0}^1 u_s v_s$, the adjoint null-vector $\bm{\phi}^\dagger(\xi)$ satisfies the adjoint equation obtained via integration by parts for functions vanishing at $\xi \to \pm\infty$:
\begin{equation}
    \mathcal{J}^\dagger \bm{\phi}^\dagger = V_{\mathrm{flat}} \partial_\xi \bm{\phi}^\dagger + \mathsf{P}(\xi)^\top \bm{\phi}^\dagger + \partial_\xi^2 \left( \mathsf{Q}(\xi)^\top \bm{\phi}^\dagger \right) = 0.
    \label{eq:app_adjoint_operator}
\end{equation}
Note that the dagger in $\bm{\phi}^\dagger$ indicates the adjoint operator with respect to this inner product, rather than the conjugate transpose of a vector $\bm{\phi}$.
According to the Fredholm alternative, a bounded perturbation solution $\bm{\rho}^{(1)}$ exists if and only if the right-hand side of Eq.~\eqref{eq:app_linear_inhomogeneous_ode} is orthogonal to the adjoint null-vector $\bm{\phi}^\dagger(\xi)$:
\begin{equation}
    \int_{-\infty}^{\infty} \bm{\phi}^\dagger(\xi) \cdot \left[ - D_{\mathrm{eff}} \partial_\xi \bm{\rho}^{(0)}(\xi) + \bm{N}(\bm{\rho}^{(0)}(\xi)) \right] d\xi = 0.
\end{equation}
Solving this orthogonality condition directly yields the integral expression for $D_{\mathrm{eff}}$ in Eq.~\eqref{eq:deff_integral_ratio}.

\subsection{Inhomogeneous Terms and Jacobian Operator}\label{app:curvature_algebraic_details}
The inhomogeneous source term $\bm{N}(\bm{\rho}^{(0)}) = (N_0, N_1)^\top$ in Eq.~\eqref{eq:app_linear_inhomogeneous_ode} arises from two distinct geometric contributions: the expansion of the Laplacian in polar coordinates ($\nabla^2 = \partial_\xi^2 + \kappa \partial_\xi$) and the curvature correction to the transition rates $w(s \rightarrow s')$ through the local potential difference.

Specifically, the local potential $U_s(\xi)$ is expanded as
\begin{equation}
    U_s(\xi) = U_s^{(0)}(\xi) + \kappa U_s^{(1)}(\xi) + \mathcal{O}(\kappa^2),
\end{equation}
where the leading and first-order contributions are given by
\begin{equation}
    U_s^{(0)}(\xi) = -z J \left[ \rho_s^{(0)}(\xi) + \frac{\sigma^2}{4} \partial_\xi^2 \rho_s^{(0)}(\xi) \right],
\end{equation}
and
\begin{equation}
    U_s^{(1)}(\xi) = -z J \left[ \rho_s^{(1)}(\xi) + \frac{\sigma^2}{4} \left( \partial_\xi^2 \rho_s^{(1)}(\xi) + \partial_\xi \rho_s^{(0)}(\xi) \right) \right],
    \label{eq:app_potential_first_order}
\end{equation}
respectively.
The continuous transition rate $w(s \rightarrow s'; \xi)$ in Eq.~\eqref{eq:transition_rate_contin} is expanded in powers of $\kappa$ as
\begin{equation}
    w(s \rightarrow s'; \xi) = w^{(0)}(s \rightarrow s'; \xi) + \kappa w^{(1)}(s \rightarrow s'; \xi) + \mathcal{O}(\kappa^2).
    \label{eq:app_w_expansion}
\end{equation}
Here, the transition rate depends on the local potential difference $\Delta U_{s, s'}(\xi) \coloneqq U_{s'}(\xi) - U_s(\xi)$, which is expanded in powers of $\kappa$ as
\begin{equation}
    \Delta U_{s, s'}(\xi) = \Delta U_{s, s'}^{(0)}(\xi) + \kappa \Delta U_{s, s'}^{(1)}(\xi) + \mathcal{O}(\kappa^2).
    \label{eq:app_delta_u_expansion}
\end{equation}
The linear perturbation in $\Delta U_{s, s'}^{(1)}$ proportional to $\bm{\rho}^{(1)}$ is incorporated into the Jacobian operator $\mathcal{J}$, whereas the remaining inhomogeneous part $w^{(1)}(s \rightarrow s'; \xi)$ depends solely on the unperturbed 1D front profile:
\begin{equation}
    w^{(1)}(s \rightarrow s'; \xi) = - \frac{\sigma^2}{4} z J w'_{s \rightarrow s'} \cdot \left[ \partial_\xi \rho_s^{(0)}(\xi) - \partial_\xi \rho_{s'}^{(0)}(\xi) \right],
\end{equation}
with the derivative of the transition rate evaluated at the unperturbed potential difference:
\begin{equation}
    w'_{s \rightarrow s'} \coloneqq \frac{\beta}{4\tau} \sech^2\left( \frac{\beta}{2} \left( \Delta U_{s, s'}^{(0)} - h_{s, s'} \right) \right).
\end{equation}
Because the probability conservation constraint $\rho_2 = 1 - \rho_0 - \rho_1$ leaves two independent components, the components $N_s$ ($s \in \{0, 1\}$) of the inhomogeneous vector $\bm{N}(\bm{\rho}^{(0)})$ are evaluated as
\begin{align}
    &N_s(\xi) \notag\\
    &= w^{(1)}(s-1 \rightarrow s; \xi) \rho_{s-1}^{(0)}(\xi) + w^{(1)}(s+1 \rightarrow s; \xi) \rho_{s+1}^{(0)}(\xi) \notag\\
    &\quad - \left[ w^{(1)}(s \rightarrow s+1; \xi) + w^{(1)}(s \rightarrow s-1; \xi) \right] \rho_s^{(0)}(\xi).
\end{align}

Due to the conservation of probability $\rho_2 = 1 - \rho_0 - \rho_1$, the components $P_{ss'}(\xi)$ ($s, s' \in \{0, 1\}$) of the coefficient matrix $\mathsf{P}(\xi)$ in the linearized Jacobian operator $\mathcal{J}$ are given by
\begin{subequations}
\label{eq:p_matrix_components}
\begin{align}
    P_{00}(\xi) &= w_{2 \rightarrow 0}^{(0)} + w_{0 \rightarrow 1}^{(0)} + w_{0 \rightarrow 2}^{(0)} \notag \\
    &\quad - 2 z J \rho_2^{(0)} w'_{2 \rightarrow 0} - z J \rho_1^{(0)} w'_{1 \rightarrow 0} \notag \\
    &\quad - z J \rho_0^{(0)} w'_{0 \rightarrow 1} - 2 z J \rho_0^{(0)} w'_{0 \rightarrow 2}, \\
    P_{01}(\xi) &= - w_{1 \rightarrow 0}^{(0)} + w_{2 \rightarrow 0}^{(0)} - z J \rho_2^{(0)} w'_{2 \rightarrow 0} \notag \\
    &\quad + z J \rho_1^{(0)} w'_{1 \rightarrow 0} + z J \rho_0^{(0)} w'_{0 \rightarrow 1} \notag \\
    &\quad - z J \rho_0^{(0)} w'_{0 \rightarrow 2}, \\
    P_{10}(\xi) &= - w_{0 \rightarrow 1}^{(0)} + w_{2 \rightarrow 1}^{(0)} + z J \rho_0^{(0)} w'_{0 \rightarrow 1} \notag \\
    &\quad - z J \rho_2^{(0)} w'_{2 \rightarrow 1} - z J \rho_1^{(0)} w'_{1 \rightarrow 2} \notag \\
    &\quad + z J \rho_1^{(0)} w'_{1 \rightarrow 0}, \\
    P_{11}(\xi) &= w_{2 \rightarrow 1}^{(0)} + w_{1 \rightarrow 2}^{(0)} + w_{1 \rightarrow 0}^{(0)} \notag \\
    &\quad - z J \rho_0^{(0)} w'_{0 \rightarrow 1} - 2 z J \rho_2^{(0)} w'_{2 \rightarrow 1} \notag \\
    &\quad - 2 z J \rho_1^{(0)} w'_{1 \rightarrow 2} - z J \rho_1^{(0)} w'_{1 \rightarrow 0},
\end{align}
\end{subequations}
where $w_{s \rightarrow s'}^{(0)} \coloneqq w^{(0)}(s \rightarrow s'; \xi)$ is the leading-order transition rate~\eqref{eq:app_w_expansion}.
The coefficient matrix $\mathsf{Q}(\xi)$ of the second-order spatial derivative is related to the potential-derivative contribution of $\mathsf{P}(\xi)$ by
\begin{equation}
    \mathsf{Q}(\xi) = \frac{\sigma^2}{4} \left( \mathsf{P}(\xi) - \mathsf{P}^{\mathrm{base}}(\xi) \right),
\end{equation}
where $\mathsf{P}^{\mathrm{base}}(\xi)$ is the matrix of bare transition rates without the functional derivatives of the potentials, whose components are given by
\begin{subequations}
\label{eq:p_base_components}
\begin{align}
    P_{00}^{\mathrm{base}} &= w_{2 \rightarrow 0}^{(0)} + w_{0 \rightarrow 1}^{(0)} + w_{0 \rightarrow 2}^{(0)}, \\
    P_{01}^{\mathrm{base}} &= -w_{1 \rightarrow 0}^{(0)} + w_{2 \rightarrow 0}^{(0)}, \\
    P_{10}^{\mathrm{base}} &= -w_{0 \rightarrow 1}^{(0)} + w_{2 \rightarrow 1}^{(0)}, \\
    P_{11}^{\mathrm{base}} &= w_{2 \rightarrow 1}^{(0)} + w_{1 \rightarrow 2}^{(0)} + w_{1 \rightarrow 0}^{(0)}.
\end{align}
\end{subequations}

%

\clearpage
\onecolumngrid

\begin{center}
    \vspace*{0.5cm}
    {\large \textbf{Supplemental Material for:}} \\[0.2cm]
    {\large \textbf{Continuum mean-field theory of spatiotemporal pattern formation in a three-state active Potts model}} \\[0.3cm]
    Yuta Tateyama,$^{1}$ Songhao Yin,$^{1,2}$ Hiroaki Ito,$^{1}$ Hiroyuki Kitahata,$^{1}$ and Hiroshi Noguchi$^{2}$ \\[0.2cm]
    \textit{$^{1}$Department of Physics, Chiba University, Chiba 263-8522, Japan} \\
    \textit{$^{2}$Institute for Solid State Physics, University of Tokyo, Kashiwa, Chiba 277-8581, Japan} \\[0.5cm]
\end{center}

\twocolumngrid

\setcounter{section}{0}
\setcounter{subsection}{0}
\setcounter{equation}{0}
\setcounter{figure}{0}
\setcounter{table}{0}
\setcounter{page}{1}
\renewcommand{\appendixname}{Section}
\renewcommand{\thesection}{S\arabic{section}}
\renewcommand{\thesubsection}{S\arabic{section}.\arabic{subsection}}
\makeatletter
\renewcommand{\p@subsection}{}
\renewcommand{\p@subsubsection}{}
\makeatother
\renewcommand{\theequation}{S\arabic{equation}}
\renewcommand{\thefigure}{S\arabic{figure}}
\renewcommand{\thetable}{S\arabic{table}}
\renewcommand{\thepage}{S\arabic{page}}
\renewcommand{\theHsection}{SM.\arabic{section}}
\renewcommand{\theHsubsection}{SM.\arabic{section}.\arabic{subsection}}
\renewcommand{\theHfigure}{SM.\arabic{figure}}
\renewcommand{\theHtable}{SM.\arabic{table}}
\renewcommand{\theHequation}{SM.\arabic{equation}}

\section{Numerical Schemes for Continuum Simulations and Bifurcation Analysis}
\label{sm:numerical_schemes}
This section details the numerical simulation methods and analysis algorithms for the continuum mean-field equations (CMFEs)~\eqref{eq:mf_potential}--\eqref{eq:transition_rate_contin} employed throughout Secs.~\ref{sec:model_continuum}--\ref{sec:interfacial_and_spiral_dynamics} in the main text.

\subsection{Homogeneous System and Bifurcation Analysis}
\label{sm:homogeneous_system}
\begin{figure*}[tbp]
    \centering
    \includegraphics[width=\textwidth]{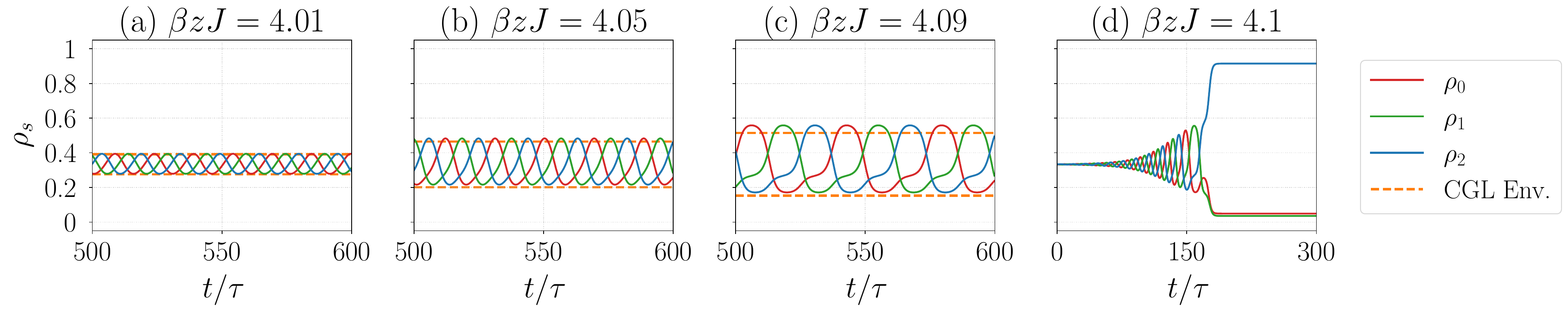}
    \caption{
      Temporal waveforms in the spatially homogeneous system of the CMFEs~\eqref{eq:mf_potential}--\eqref{eq:transition_rate_contin} for $\zeta=0.5$.
      (a)--(c) Periodic limit cycles at (a) $\beta z J = 4.01$, (b) $\beta z J = 4.05$, and (c) $\beta z J = 4.09$ ($500 \leq t/\tau \leq 600$).
      (d) Relaxation trajectory towards an ordered fixed point at $\beta z J = 4.1$ ($0 \leq t/\tau \leq 300$).
      The orange dashed lines in (a) represent the oscillation envelope predicted by the weakly nonlinear CGL analysis.
      The trajectories are calculated starting from initial states with a small perturbation added to the unstable homogeneous disordered state.
    }
    \label{fig:zeta-0_5-rho_time}
\end{figure*}

This section provides the numerical procedures for steady-state continuation and periodic orbit tracking of the spatially homogeneous system presented in Sec.~\ref{sec:homogeneous_limit_dynamics} of the main text, which are used to construct the bifurcation diagrams and phase portraits shown in Fig.~\ref{fig:bifurcation_diagrams_combined}.
The fixed points were determined by numerically solving the steady-state algebraic equations using a Newton--Raphson algorithm, and their linear stabilities were evaluated via the eigenvalue analysis of the Jacobian matrix of the spatially homogeneous ordinary differential equation (ODE) system~\eqref{eq:mf_potential}--\eqref{eq:transition_rate_contin}.
The bifurcation curves of the limit cycles in Figs.~\ref{fig:bifurcation_diagrams_combined}(a) and \ref{fig:bifurcation_diagrams_combined}(d) display the period-averaged order parameter $\overline{R}_1 \coloneqq (1 / T) \int_0^T R_1(t)\, dt$.
The period $T$ of each limit cycle was determined by identifying successive crossings of the Poincar\'e section $\rho_1 = \rho_2$ with $\dot{\rho}_1 > \dot{\rho}_2$, and the temporal average was evaluated by numerically integrating $R_1(t)$ over one period using the trapezoidal rule after discarding initial transients.
In the subcritical regime ($\zeta = 0.7$), the branch of unstable limit cycles was tracked by integrating the time-reversed dynamical system.
These numerical computations were performed using the root-finding and adaptive ODE integration routines in the SciPy library~\cite{virtanen2020scipy}.
The $\mathbb{Z}_3$-symmetric heteroclinic bifurcation point $\beta z J_{\mathrm{het}}$ was identified by sweeping the coupling parameter $\beta z J$ with adaptive steps and monitoring the divergence of the limit-cycle period.
Figure~\ref{fig:zeta-0_5-rho_time} illustrates representative temporal waveforms of the limit cycle for $\zeta=0.5$ at various coupling strengths $\beta z J$.
At the onset of the Hopf bifurcation ($\beta z J = 4.01$), the waveform displays small-amplitude nearly sinusoidal oscillations whose amplitude envelope agrees with the prediction from the weakly nonlinear complex Ginzburg--Landau (CGL) analysis in Sec.~\ref{sec:cgl} of the main text.
As $\beta z J$ increases, higher-harmonic contributions generated by the $(A^*)^2$ term progressively distort the waveform, and the trajectory eventually collides with the saddles at the $\mathbb{Z}_3$-symmetric heteroclinic bifurcation point $\beta z J_{\mathrm{het}} \simeq 4.092$.

\subsection{One-Dimensional Nonlinear Boundary Value Problem}\label{sm:bvp_numerical_scheme}
This section provides the numerical algorithm used to solve the one-dimensional (1D) nonlinear boundary value problem (BVP) formulated in Sec.~\ref{sec:traveling_front}, which yields the traveling front profile shown in Fig.~\ref{fig:bvp_summary}(a) and the propagation velocity $V_{\mathrm{flat}}$ displayed as solid curves in Figs.~\ref{fig:bvp_summary}(b) and \ref{fig:circular_front_sweep_summary}(a).
To obtain the steady traveling front profile and the propagation velocity $V$, we seek a traveling front solution of the form $\rho_s(x, t) = \rho_s(\xi)$ in the comoving frame $\xi \coloneqq x - Vt$.
Under the constraint $\sum_{s=0}^2 \rho_s(\xi) = 1$, the traveling front equations are given by Eq.~\eqref{eq:traveling_front_ode} for $s \in \{0, 1\}$, where the transition rates $w(s \rightarrow s')$ are given by Eq.~\eqref{eq:transition_rate_contin} with the 1D local potential:
\begin{equation}
    U_s(\xi) = -zJ \left( \rho_s + \frac{\sigma^2}{4} \frac{d^2\rho_s}{d\xi^2} \right).
    \label{eq:app_bvp_potential}
\end{equation}
Because the second derivatives $d^2\rho_s/d\xi^2$ are implicitly enclosed inside the nonlinear hyperbolic tangent functions of the Glauber-type rates, Eq.~\eqref{eq:traveling_front_ode} cannot be rearranged into an explicit form.

To address this implicit structure, we employ a nested numerical scheme combining a global collocation method and a local root-finding algorithm.
First, to handle the boundary conditions and establish the coordinate origin, the physical domain $\xi \in [-L, L]$ is decomposed into two symmetric intervals: the left half $[-L, 0]$ and the right half $[0, L]$.
Both intervals are mapped onto a normalized coordinate $z \in [0, 1]$.
The boundary conditions at the outer boundaries are set to the bulk ordered states: $\rho_s(-L) = \rho_{s,\mathrm{L}}$ (dominated by state $s = 0$) and $\rho_s(L) = \rho_{s,\mathrm{R}}$ (dominated by state $s = 1$).
At the junction $\xi = 0$, we enforce $C^1$ continuity: $\rho_s(-0) = \rho_s(+0)$ and $\rho'_s(-0) = \rho'_s(+0)$ for $s \in \{0, 1\}$, where the prime denotes the derivative with respect to $\xi$.
The translational invariance is broken by introducing the front pinning condition at the junction:
\begin{equation}
    \rho_0(0) = \frac{\rho_{0,\mathrm{L}} + \rho_{0,\mathrm{R}}}{2}.
\end{equation}
This additional constraint uniquely determines the eigenvalue $V$.
The resulting BVP system is solved using the fourth-order collocation method (as implemented in the SciPy library).

Second, during the residual evaluation of the global BVP, the implicit algebraic equations in Eq.~\eqref{eq:traveling_front_ode} are solved numerically for the second derivatives at each grid point using a local two-dimensional (2D) Newton--Raphson algorithm.
To ensure stable convergence, the second-order derivatives from the previous collocation iteration are used as the initial guess for the Newton--Raphson solver.

To compute the traveling front profiles over a wide range of the coupling strength $\beta z J$, we employ a sequential parameter continuation method from $\beta z J = 3.9$ to $10$ with adaptive step size control.

\subsection{1D Front Simulations and Interface Width}\label{sm:1d_pde_scheme}
This section details the numerical integration scheme for direct numerical simulations of the 1D CMFEs and the mathematical definition of the interface width $W$ used in Sec.~\ref{sec:traveling_front} and plotted in Fig.~\ref{fig:bvp_summary}(c).
For the 1D CMFE simulation based on Eqs.~\eqref{eq:mf_potential}--\eqref{eq:transition_rate_contin}, the spatial grid size was set to $\Delta x/\sigma = 0.01$ with system size $L/\sigma = 60$ ($N = 6001$ grid points).
The 1D Laplace operator $\partial_x^2$ was approximated with second-order accuracy using the central difference scheme with fixed Dirichlet boundary conditions at both ends.
Specifically, the variables $\rho_s$ at the left boundary $x = -L/2$ were fixed to the ordered state $s = 0$.
Similarly, those at the right boundary $x = L/2$ were fixed to the ordered state $s = 1$ (for details of the fixed-point calculation, see Sec.~\ref{sm:homogeneous_system}).
The time evolution of the resulting stiff ODE system was performed using the fourth-order explicit singly diagonal implicit Runge--Kutta (ESDIRK) method with adaptive time-stepping~\cite{rackauckas2017differentialequations}.

In the numerical analysis, the interface width $W$ of the traveling front was quantified as the maximum gradient width, defined by
\begin{equation}
    W \coloneqq \frac{\rho_{0, \mathrm{L}} - \rho_{0, \mathrm{R}}}{\max_\xi | \partial_\xi \rho_0 |},
    \label{eq:interface_width_def}
\end{equation}
where $\rho_{0, \mathrm{L}} - \rho_{0, \mathrm{R}}$ represents the total variation of the major component $\rho_0$ across the interface, and $\max_\xi | \partial_\xi \rho_0 |$ is the maximum slope of the steady front profile $\rho_0(\xi)$.

\subsection{Numerical Verification of Traveling Front Velocity}\label{sm:pde_verification}
\begin{figure}[tbp]
    \centering
    \includegraphics[width=\columnwidth]{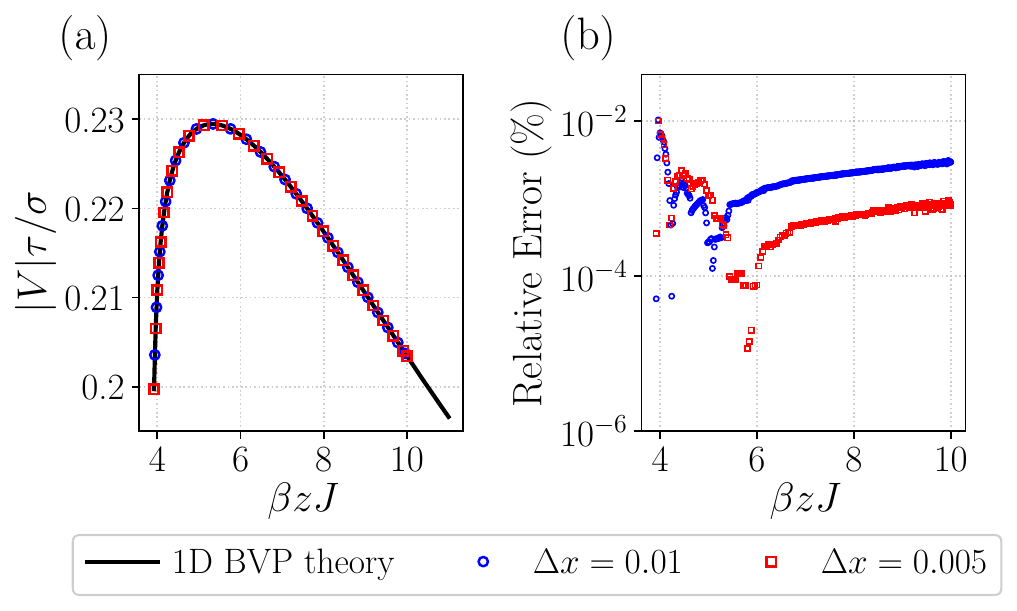}
    \caption{
        Quantitative comparison of the dimensionless 1D steady traveling front velocity $|V|\tau/\sigma$ obtained from the 1D BVP analysis and 1D direct CMFE simulations~\eqref{eq:mf_potential}--\eqref{eq:transition_rate_contin} for $\zeta = 0.5$.
        (a) Dimensionless velocity $|V|\tau/\sigma$ as a function of the coupling strength $\beta z J$.
        The black solid line denotes the 1D BVP theory.
        Markers denote 1D CMFE simulations for grid spacings $\Delta x/\sigma = 0.01$ (blue circles) and $\Delta x/\sigma = 0.005$ (red squares).
        (b) Relative error of the 1D CMFE velocity with respect to the 1D BVP theory.
    }
    \label{fig:pde_sweep_comparison_errors}
\end{figure}

This section provides direct numerical simulations of the 1D CMFEs~\eqref{eq:mf_potential}--\eqref{eq:transition_rate_contin} to confirm that the steady traveling front profiles and propagation velocities obtained from the 1D BVP~\eqref{eq:traveling_front_ode} are physically realized and stable against spatial perturbations, corresponding to the velocity comparison and relative error plots shown in Fig.~\ref{fig:pde_sweep_comparison_errors}.

The system size is set to $L / \sigma = 60$, which is sufficiently wider than the interface width, with a grid spacing of $\Delta x / \sigma = 0.01$, and the boundary conditions at both ends are fixed to the ordered states: state $s = 0$ at the left boundary $x \to -L/2$ and state $s = 1$ at the right boundary $x \to L/2$ (see Sec.~\ref{sm:homogeneous_system}).
As the initial condition, we impose a $\tanh$-type interface profile smoothly connecting both ordered states, and evolve the system in time using an implicit Runge--Kutta method with adaptive time-stepping (for the numerical scheme, see Sec.~\ref{sm:1d_pde_scheme}).

To evaluate the propagation velocity $V_{\mathrm{CMFE}}$, we track the front position $x_{\mathrm{f}}(t)$ defined as the location where the local occupancy $\rho_0(x, t)$ crosses the midpoint between the left and right bulk values $\rho_{0,\mathrm{mid}} = (\rho_{0,\mathrm{L}} + \rho_{0,\mathrm{R}}) / 2$, and determine the slope by applying a linear regression $x_{\mathrm{f}}(t) = x_0 + V_{\mathrm{CMFE}} t$ over a dimensionless time window $t/\tau \in [25, 80]$ that excludes initial transient relaxation.

\begin{figure}[tbp]
    \centering
    \includegraphics[width=\columnwidth]{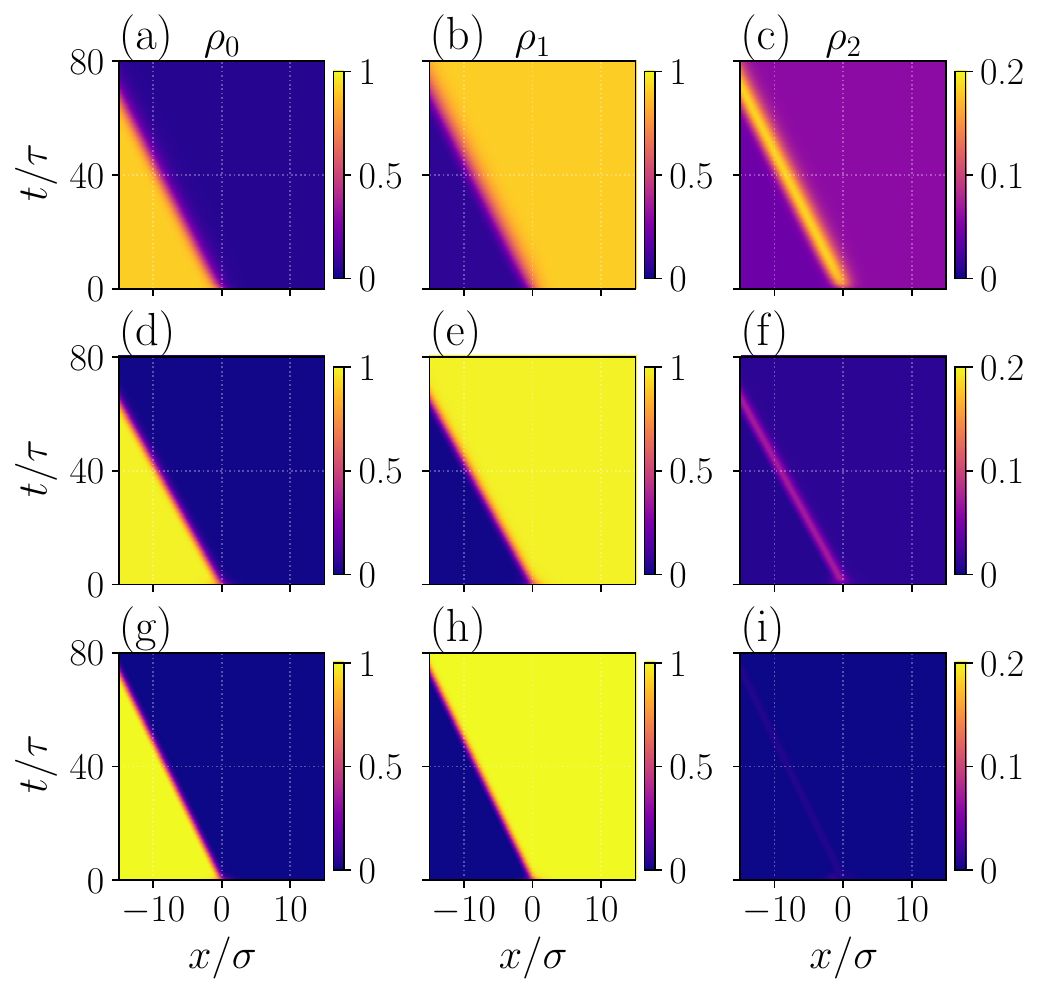}
    \caption{Spatiotemporal plots of the traveling front obtained from direct numerical simulations of the 1D CMFEs~\eqref{eq:mf_potential}--\eqref{eq:transition_rate_contin} for coupling strengths (a)--(c) $\beta z J = 4$, (d)--(f) $\beta z J = 5.32$, and (g)--(i) $\beta z J = 10$.
    The left, middle, and right columns show the spatiotemporal distributions of $\rho_0(x,t)$, $\rho_1(x,t)$, and $\rho_2(x,t)$, respectively.}
    \label{fig:pde_simulation_verification}
\end{figure}

Figure~\ref{fig:pde_simulation_verification} shows spatiotemporal plots of $\rho_0(x,t)$, $\rho_1(x,t)$, and $\rho_2(x,t)$ obtained for three representative values of the coupling strength: $\beta z J = 4$, $5.32$, and $10$.
The interface propagates to the left at a constant velocity while stably maintaining its shape.
In particular, the results consistently show that the intermediate state $\rho_2$ transiently forms a sharp local peak at the interface and propagates stably during the transition from state $0$ to state $1$.

Furthermore, Fig.~\ref{fig:pde_sweep_comparison_errors} shows a quantitative comparison of the dimensionless 1D front propagation velocity $|V|\tau/\sigma$ obtained from the 1D BVP analysis and 1D direct CMFE simulations (with $\Delta x / \sigma = 0.01$ and $\Delta x / \sigma = 0.005$) for $4 \leq \beta z J \leq 10$, as well as their relative errors.
As shown in Fig.~\ref{fig:pde_sweep_comparison_errors}(a), the two results agree well over the entire range of $\beta z J$, with the relative error below approximately $0.01\%$ (and below $0.005\%$ in most regions) as shown in Fig.~\ref{fig:pde_sweep_comparison_errors}(b).

\subsection{Adjoint Null-Vector and Effective Diffusion Coefficient}\label{sm:adjoint_numerical_scheme}
This section describes the numerical algorithm used to determine the adjoint null-vector $\bm{\phi}^\dagger(\xi)$ and evaluate the effective diffusion coefficient $D_{\mathrm{eff}}$ presented in Sec.~\ref{sec:circular_front}, which yields the theoretical prediction shown as the solid curve in Fig.~\ref{fig:circular_front_sweep_summary}(b).
We solve the adjoint eigenvalue problem of the linearized Jacobian operator $\mathcal{J}$ in the comoving frame.
Using the probability conservation constraint $\sum_{s=0}^2 \rho_s(\xi) = 1$, we reduce the system to two independent variables $\rho_0(\xi)$ and $\rho_1(\xi)$.
The linearized Jacobian operator $\mathcal{J}$ acting on the perturbations $(\delta\rho_0, \delta\rho_1)^\top$ is discretized on a spatial grid of $N = 300$ points.
The first- and second-order spatial derivatives are approximated using second-order finite-difference stencils.
The discretized adjoint operator matrix $\mathcal{J}^\dagger$ is constructed, and its eigenvalue problem is solved using a dense eigensolver to identify the eigenvector corresponding to the eigenvalue closest to zero, which represents the adjoint null-vector $\bm{\phi}^\dagger(\xi) = (\phi^\dagger_0(\xi), \phi^\dagger_1(\xi))^\top$.

\begin{figure}[tbp]
    \centering
    \includegraphics[width=\columnwidth]{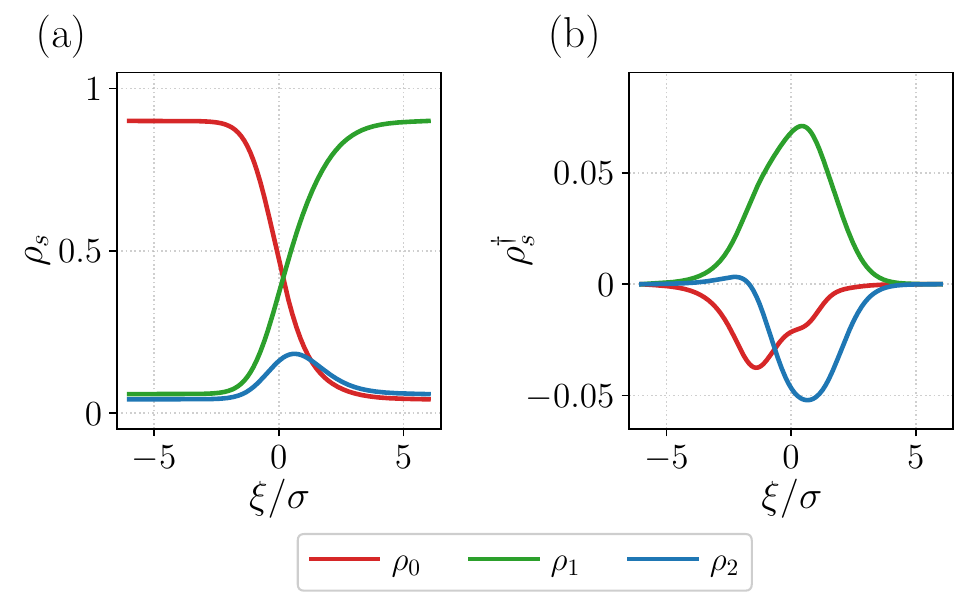}
    \caption{1D traveling front profiles and the corresponding adjoint null-vector components at $\beta z J = 4$.
    (a) Stationary density profiles $\rho_s(\xi)$ of the three states $s \in \{0, 1, 2\}$ in the comoving frame.
    (b) Components of the symmetric adjoint null-vector $\rho^\dagger_s(\xi)$, showing the sensitivity of the interface position to local perturbations in each state density.
    All components in (b) are strongly localized at the interface region.}
    \label{fig:adjoint_mode_profile}
\end{figure}

To restore the physical symmetry among the three states, we introduce the symmetric gauge condition $\sum_{s=0}^2 \rho^\dagger_s(\xi) = 0$ for the adjoint components under the conservation law $\sum_{s=0}^2 \delta\rho_s(\xi) = 0$.
Because the projection satisfies $\sum_{s=0}^2 \rho^\dagger_s \delta\rho_s = \phi^\dagger_0 \delta\rho_0 + \phi^\dagger_1 \delta\rho_1$ for any constrained variations with $\delta\rho_2 = -\delta\rho_0 - \delta\rho_1$, equating coefficients yields $\rho^\dagger_0 - \rho^\dagger_2 = \phi^\dagger_0$ and $\rho^\dagger_1 - \rho^\dagger_2 = \phi^\dagger_1$.
Combining these with the symmetric gauge condition uniquely yields the components:
\begin{subequations}
    \begin{align}
        \rho^\dagger_0(\xi) &= \frac{2\phi^\dagger_0(\xi) - \phi^\dagger_1(\xi)}{3}, \\
        \rho^\dagger_1(\xi) &= \frac{2\phi^\dagger_1(\xi) - \phi^\dagger_0(\xi)}{3}, \\
        \rho^\dagger_2(\xi) &= -\frac{\phi^\dagger_0(\xi) + \phi^\dagger_1(\xi)}{3}.
    \end{align}
\end{subequations}
Figure~\ref{fig:adjoint_mode_profile} shows the density profiles $\rho_s(\xi)$ and the corresponding symmetric adjoint null-vector components $\rho^\dagger_s(\xi)$ computed for the representative coupling strength $\beta z J = 4$.
Both components are highly localized at the interface transition region, validating the local eikonal approximation.
Using the localized adjoint profile, the theoretical effective diffusion coefficient $D_{\mathrm{eff}}$ is evaluated by numerically calculating the projection integrals in Eq.~\eqref{eq:deff_integral_ratio} using trapezoidal integration.

\subsection{2D Circular Front Simulations}
\label{sm:circular_front_scheme}
This section outlines the numerical simulation scheme for circular domain fronts and the dynamic interface tracking algorithm used in Sec.~\ref{sec:circular_front}, which generates the circular front trajectories and simulation points shown in Figs.~\ref{fig:circular_front_individual} and \ref{fig:circular_front_sweep_summary}.
The 2D CMFEs~\eqref{eq:mf_potential}--\eqref{eq:transition_rate_contin} were numerically integrated on a square grid with periodic boundary conditions.
The system size was set to $L/\sigma = 40$ and the grid spacing was set to $\Delta x/\sigma = 0.2$, resulting in $200 \times 200$ spatial grid points.
The spatial Laplacian $\nabla^2$ was approximated using the standard 5-point finite-difference stencil.
For time integration, we employed the explicit second-order Heun's method with a fixed time step size $\Delta t / \tau = 0.8 \Delta x^2 / (\beta z J \sigma^2)$.
To initialize the circular domain of state $s$ with initial radius $R_0$, the spatial profiles were set as:
\begin{align}
    \rho_{s'}(\bm{x}, 0) &= \rho_{s', \mathrm{out}} \notag\\
    &\quad + \frac{\rho_{s', \mathrm{in}} - \rho_{s', \mathrm{out}}}{2} \left[ 1 - \tanh \left( \frac{k(r_{\mathrm{d}} - R_0)}{\sigma}\right) \right],
\end{align}
where $r_{\mathrm{d}} \coloneqq \sqrt{(x - x_{\mathrm{c}})^2 + (y - y_{\mathrm{c}})^2}$ is the distance from the center $(x_{\mathrm{c}}, y_{\mathrm{c}})$, $k = 1.5$ is an empirical parameter chosen to set a smooth initial profile, and the interior and exterior densities, $\bm{\rho}_{\mathrm{in}}$ and $\bm{\rho}_{\mathrm{out}}$, correspond to the ordered states obtained in Sec.~\ref{sm:homogeneous_system}.
For the convex front configuration, the domain interior is set to state $s = 1$ ($\bm{\rho}_{\mathrm{in}} = \bm{\rho}_{\mathrm{R}}$) and the background to state $s = 0$ ($\bm{\rho}_{\mathrm{out}} = \bm{\rho}_{\mathrm{L}}$) with initial radius $R_0/\sigma = 5$.
Because state $s = 1$ invades state $s = 0$, the circular front advances outward with $dR/dt > 0$, presenting an interface that is convex with respect to the forward invasion direction.
Conversely, for the concave front configuration, the domain interior is set to state $s = 0$ ($\bm{\rho}_{\mathrm{in}} = \bm{\rho}_{\mathrm{L}}$) and the background to state $s = 1$ ($\bm{\rho}_{\mathrm{out}} = \bm{\rho}_{\mathrm{R}}$) with initial radius $R_0/\sigma = 15$.
In this case, the background of state $s = 1$ invades inward into the domain interior of state $s = 0$ with $dR/dt < 0$, presenting an interface that is concave with respect to the forward invasion direction.

To dynamically track the radius $R(t)$ of the circular front, we evaluate the angular-averaged position of the interface.
Specifically, we sample the density profile $\rho_0$ along eight equally spaced radial directions $\theta_j = j \pi / 4$ ($j = 0, \dots, 7$) from the domain center via bilinear interpolation on the 2D grid (with radial resolution $\Delta r/\sigma \simeq 0.02$).
Along each direction $\theta_j$, the crossing radius $r_{\mathrm{c}}(\theta_j, t)$ where $\rho_0$ reaches the threshold value $\rho_{0, \mathrm{mid}} \coloneqq (\rho_{0, \mathrm{L}} + \rho_{0, \mathrm{R}})/2$ is extracted by linear interpolation.
The front radius $R(t)$ is then obtained by averaging $r_{\mathrm{c}}(\theta_j, t)$ over these eight directions.

For the eikonal fitting, the tracked trajectory $R(t)$ is restricted to the interval $t/\tau > 5$ and $R(t)/\sigma > 5$ to eliminate boundary effects and transient relaxation.
The raw $R(t)$ data are smoothed using a moving average filter with a window size of 15 steps.
The local normal propagation velocity along the forward invasion direction is evaluated as $V_{\mathrm{n}}(t) = dR/dt$ for the convex front and $V_{\mathrm{n}}(t) = -dR/dt$ for the concave front using central differences of the smoothed $R(t)$.
Correspondingly, the signed local curvature is defined as $\kappa(t) = 1/R(t)$ for the convex front and $\kappa(t) = -1/R(t)$ for the concave front.
Finally, the flat-interface propagation velocity $V_{\mathrm{flat}}$ and the effective diffusion coefficient $D_{\mathrm{eff}}$ are extracted by a linear regression to Eq.~\eqref{eq:eikonal} within the small-curvature regime $|\kappa|\sigma \leq 0.15$.

\begin{figure}[tbp]
    \centering
    \includegraphics[width=\columnwidth]{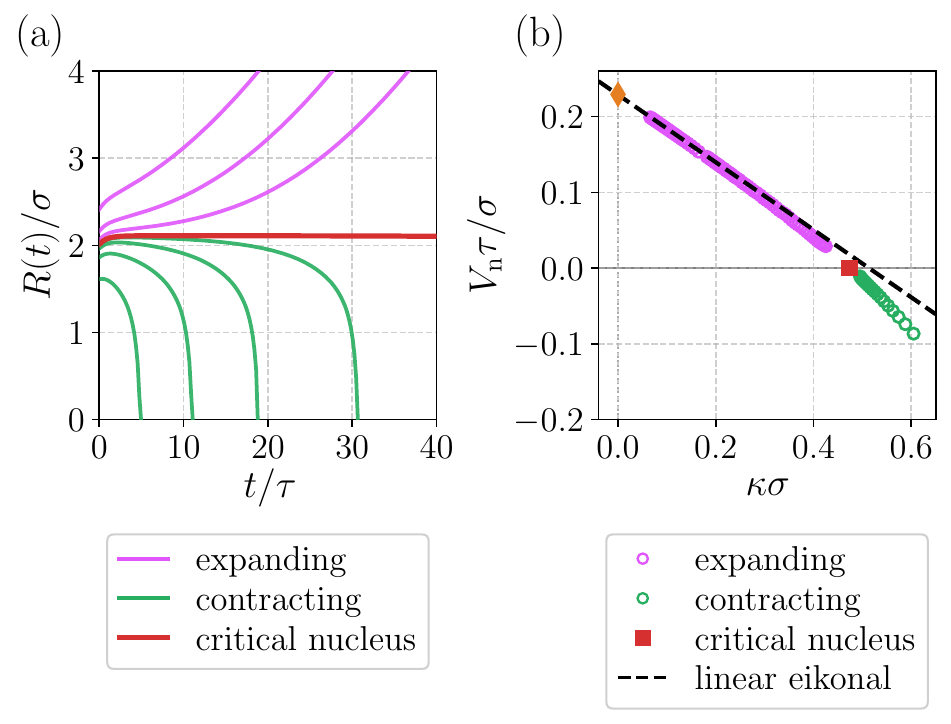}
    \caption{
        Critical nucleus dynamics and curvature-induced velocity reversal at $\beta z J = 5.5$.
        (a) Time evolution of the dimensionless domain radius $R(t)/\sigma$ for circular domains initialized near the critical nucleus radius $R_{\mathrm{c}}$.
        The red curve denotes the stationary critical nucleus trajectory ($R_{\mathrm{c}} \simeq 2.11\sigma$), separating spontaneous expansion ($R_0 > R_{\mathrm{c}}$, magenta curves) from spontaneous collapse ($R_0 < R_{\mathrm{c}}$, green curves).
        (b) Dimensionless normal propagation velocity $V_{\mathrm{n}}\tau/\sigma$ as a function of the dimensionless curvature $\kappa\sigma$.
        Magenta and green open circles denote 2D CMFE simulation data for expanding and contracting fronts, respectively.
        The red square marks the stationary critical nucleus state $V_{\mathrm{n}} = 0$.
        The orange diamond indicates the dimensionless flat-interface velocity $V_{\mathrm{flat}}\tau/\sigma$ obtained from the 1D BVP analysis at $\kappa = 0$, and the black dashed line represents the linear eikonal prediction $V_{\mathrm{n}} = V_{\mathrm{flat}} - D_{\mathrm{eff}}\kappa$.
    }
    \label{fig:sm_critical_nucleus}
\end{figure}

Beyond the small-curvature regime, extrapolating the linear eikonal equation~\eqref{eq:eikonal} to $V_{\mathrm{n}} = 0$ provides a first-order estimate for the critical nucleus radius, $R_{\mathrm{c}} \coloneqq D_{\mathrm{eff}} / V_{\mathrm{flat}}$.
To examine the critical nucleus dynamics and the nonlinear curvature effects beyond the small-curvature regime, we performed 2D CMFE simulations for circular domains with initial radii $R_0$ finely varied around $R_{\mathrm{c}}$ at $\beta z J = 5.5$, as shown in Fig.~\ref{fig:sm_critical_nucleus}.
As shown in Fig.~\ref{fig:sm_critical_nucleus}(a), a stationary critical nucleus trajectory ($R_{\mathrm{c}} \simeq 2.11\sigma$, red curve) acts as a separatrix between spontaneous expansion ($R_0 > R_{\mathrm{c}}$, magenta curves) and spontaneous collapse ($R_0 < R_{\mathrm{c}}$, green curves) in the convex configuration.
Figure~\ref{fig:sm_critical_nucleus}(b) demonstrates that while the front velocity quantitatively agrees with the linear eikonal prediction (black dashed line) at small curvatures, it progressively deviates downward as the curvature increases.
Consequently, the stationary state $V_{\mathrm{n}} = 0$ (red square) emerges at a curvature ($\kappa \simeq 0.47\sigma^{-1}$) slightly smaller than the linear estimate ($\kappa \simeq 0.51\sigma^{-1}$), beyond which the velocity drops steeply into the negative regime ($V_{\mathrm{n}} < 0$, green open circles), quantitatively capturing the nonlinear curvature-induced collapse.

\subsection{2D Spiral Wave Simulations and Absorbing Boundaries}
\label{sm:absorbing_boundary}
This section describes the numerical integration methods and absorbing boundary conditions applied in the 2D CMFE simulations of spiral waves in Sec.~\ref{sec:spiral_waves}, which produce the spiral wave profiles and simulation data points shown in Figs.~\ref{fig:spiral_fitting} and \ref{fig:sweep_J_dependencies}.
We numerically integrated the 2D CMFEs~\eqref{eq:mf_potential}--\eqref{eq:transition_rate_contin} on a square grid with spatial step size $\Delta x/\sigma = 0.35$ and system size $L/\sigma = 600$.
The 2D Laplace operator $\nabla^2$ was discretized using the standard 5-point finite-difference stencil.
For temporal integration, we employed the explicit second-order Heun's method with a time step satisfying the same numerical stability condition as in Sec.~\ref{sm:circular_front_scheme}.

To maintain a stable isolated spiral core by preventing the reflection and interference of waves propagating from the spiral arms at the boundaries, we introduced a circular absorbing buffer zone outside a central circular domain of radius $R_{\mathrm{domain}} = L/2$ centered at $(x_{\mathrm{c}}, y_{\mathrm{c}}) = (L/2, L/2)$.
Immediately after each time step, for grid points within the buffer layer $R_{\mathrm{domain}} - w_{\mathrm{buf}} \le r < R_{\mathrm{domain}}$ at radial distance $r = \sqrt{(x - x_{\mathrm{c}})^2 + (y - y_{\mathrm{c}})^2}$, the density $\rho_s$ of each component is damped towards the isotropic disordered state $\rho_s = 1/3$:
\begin{equation}
    \rho_s(\bm{x}, t) \leftarrow (1 - \lambda(d)) \rho_s(\bm{x}, t) + \frac{1}{3} \lambda(d),
    \label{eq:absorbing_boundary_damping}
\end{equation}
where $d \coloneqq R_{\mathrm{domain}} - r$ is the distance from the outer domain boundary, and the damping factor increases quadratically towards the boundary as $\lambda(d) \coloneqq \lambda_0 (1 - d/w_{\mathrm{buf}})^2$.
In the outer region $r \ge R_{\mathrm{domain}}$, the densities are pinned to the disordered state $\rho_s = 1/3$.
We set $w_{\mathrm{buf}}/\sigma = 8.4$ (corresponding to 24 grid points) and $\lambda_0 = 0.15$.

To determine the steady rotational angular velocity $\Omega$ and pitch parameter $b$ of the spiral wave, we analyze the spatiotemporal evolution of the isophase line defined by $\arg\left[ \sum_{s=0}^2 \rho_s(\bm{x}, t) e^{2\pi i s/3} \right] = 0$.
The core center $(x_{\mathrm{c}}, y_{\mathrm{c}})$ is identified as the local minimum of the amplitude $|\sum_{s=0}^2 \rho_s(\bm{x}, t) e^{2\pi i s/3}|$.
To measure the rotational angular velocity $\Omega$, we track the angular position $\theta(t)$ where the isophase line crosses a circular probe path of radius $R_{\mathrm{probe}} = 0.15 L$ ($R_{\mathrm{probe}}/\sigma = 90$) centered at the core.
After unwrapping the $2\pi$ phase jumps, we determine $\Omega$ by a linear regression $\theta(t) = \theta_0 + \Omega t$ over the steady rotation regime, excluding initial transient relaxation.
To extract the pitch parameter $b$, we trace the spatial coordinates $(r, \theta)$ of the isophase line radially outward from the core over the intermediate radial interval $r/\sigma \in [20, 200]$ that avoids both core singularity and boundary damping effects.
A linear regression of the unwrapped trajectory to the Archimedean spiral formula $r(\theta) = a + b\theta$ uniquely yields the pitch parameter $b$.

\section{Finite-Core Boundary Value Problem in Keener's Eikonal Theory}
\label{sm:finite_core_correction}
This section summarizes the numerical evaluation of the rotational angular-velocity eigenvalue in Keener's eikonal theory~\cite{keener1986geometrical, tyson1988singular} with a finite-core boundary condition, utilized in Sec.~\ref{sec:spiral_waves} to evaluate Eq.~\eqref{eq:spiral_scaling_prediction} and shown as solid curves in Fig.~\ref{fig:sweep_J_dependencies}.

As formulated in Sec.~\ref{sec:spiral_waves}, the steady profile of a rotating spiral arm is described by the parameter-free ODE~\eqref{eq:spiral_eikonal_ode} for the tangent angle $\chi(\tilde{r})$ in terms of the dimensionless radius $\tilde{r} \coloneqq r V_{\mathrm{flat}} / D_{\mathrm{eff}}$.
In the far field where the interface curvature vanishes, the wave front asymptotically approaches an Archimedean spiral with a uniform wavelength, imposing the far-field boundary condition $\lim_{\tilde{r} \to \infty} \chi(\tilde{r}) = \pi/2$~\cite{keener1986geometrical, tyson1988singular}.
When a spiral wave in excitable media rotates around an impermeable core of radius $r_0$, the front meets the core boundary orthogonally, imposing the zero-flux boundary condition $\chi(\tilde{r}_0) = 0$ at the dimensionless core radius $\tilde{r}_0 \coloneqq r_0 V_{\mathrm{flat}} / D_{\mathrm{eff}}$~\cite{keener1986geometrical, tyson1988singular}.
In the three-state active Potts model, the spiral tip forms a triple junction where the three domain walls meet.
Because each domain wall has a finite intrinsic interface width $W$ shown in Fig.~\ref{fig:bvp_summary}(c) and defined in Eq.~\eqref{eq:interface_width_def}, the sharp interface description breaks down within a core region of radius comparable to the interface thickness ($r \lesssim W$), where the three states overlap and form an isotropic disordered state.
We therefore approximate the effective core boundary as $r_0 \simeq W$, imposing $\chi(\tilde{r}_0) = 0$ with $\tilde{r}_0 = W V_{\mathrm{flat}} / D_{\mathrm{eff}}$.

\begin{figure}[tbp]
    \centering
    \includegraphics[width=\columnwidth]{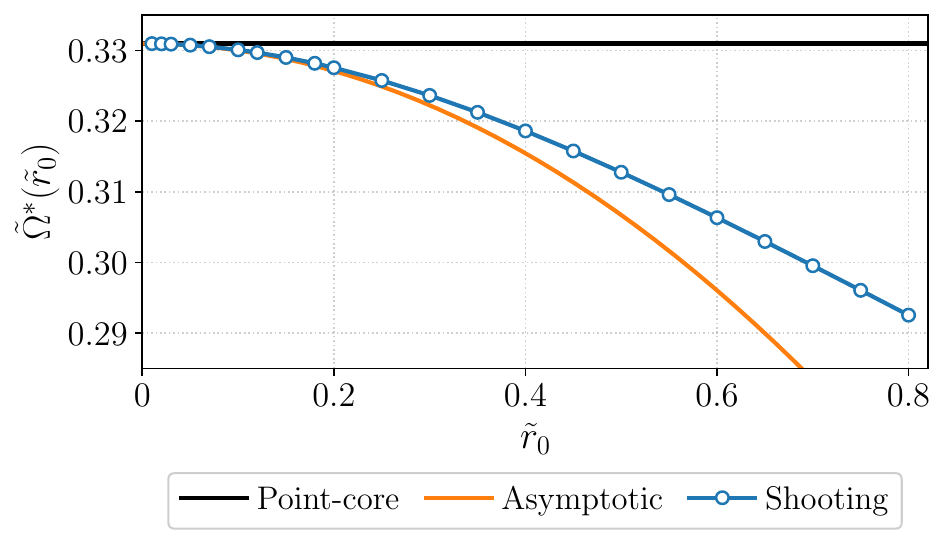}
    \caption{
        Dimensionless rotational angular-velocity eigenvalue $\tilde{\Omega}^*(\tilde{r}_0)$ as a function of the dimensionless core radius $\tilde{r}_0 = W V_{\mathrm{flat}} / D_{\mathrm{eff}}$ in Keener's eikonal theory.
        The black solid line shows the point-core limit $\tilde{\Omega}^*(0) \simeq 0.330958$.
        The orange solid line represents Keener's second-order asymptotic expansion $\tilde{\Omega}^*(\tilde{r}_0) \simeq 0.330958 - 0.097 \tilde{r}_0^2$~\cite{keener1986geometrical, tyson1988singular}.
        The blue solid line with markers denotes the eigenvalue determined by the numerical shooting method, used for the theoretical curves in Fig.~\ref{fig:sweep_J_dependencies}.
    }
    \label{fig:keener_eigenvalue_vs_core_radius}
\end{figure}

For small cores ($\tilde{r}_0 \ll 1$), Keener derived the asymptotic expansion $\tilde{\Omega}^*(\tilde{r}_0) \simeq 0.330958 - 0.097 \tilde{r}_0^2$~\cite{keener1986geometrical, tyson1988singular}.
Across the investigated coupling range ($4 \leq \beta z J \leq 10$), the dimensionless core radius reaches $\tilde{r}_0 \sim 0.3$--$0.8$, where higher-order corrections become noticeable.
We therefore compute the universal eigenvalue $\tilde{\Omega}^*(\tilde{r}_0)$ numerically using a shooting method.
Specifically, integrating Eq.~\eqref{eq:spiral_eikonal_ode} outward from $\chi(\tilde{r}_0) = 0$ with an adaptive Runge--Kutta scheme, we determine $\tilde{\Omega}^*(\tilde{r}_0)$ via bisection to satisfy the far-field boundary condition $\lim_{\tilde{r} \to \infty} \chi(\tilde{r}) = \pi/2$.
Figure~\ref{fig:keener_eigenvalue_vs_core_radius} shows the obtained eigenvalue $\tilde{\Omega}^*(\tilde{r}_0)$ alongside the point-core limit $\tilde{\Omega}^*(0) \simeq 0.330958$ and the asymptotic expansion.
While the idealized point-core limit ($\tilde{r}_0 = 0$) systematically overestimates the rotational angular velocity, incorporating the finite-core eigenvalue $\tilde{\Omega}^*(\tilde{r}_0)$ at $\tilde{r}_0 = W V_{\mathrm{flat}} / D_{\mathrm{eff}}$ accounts for this deviation and quantitatively reproduces the 2D CMFE simulation results shown in Fig.~\ref{fig:sweep_J_dependencies}.

\begin{figure}[tbp]
    \centering
    \includegraphics[width=\columnwidth]{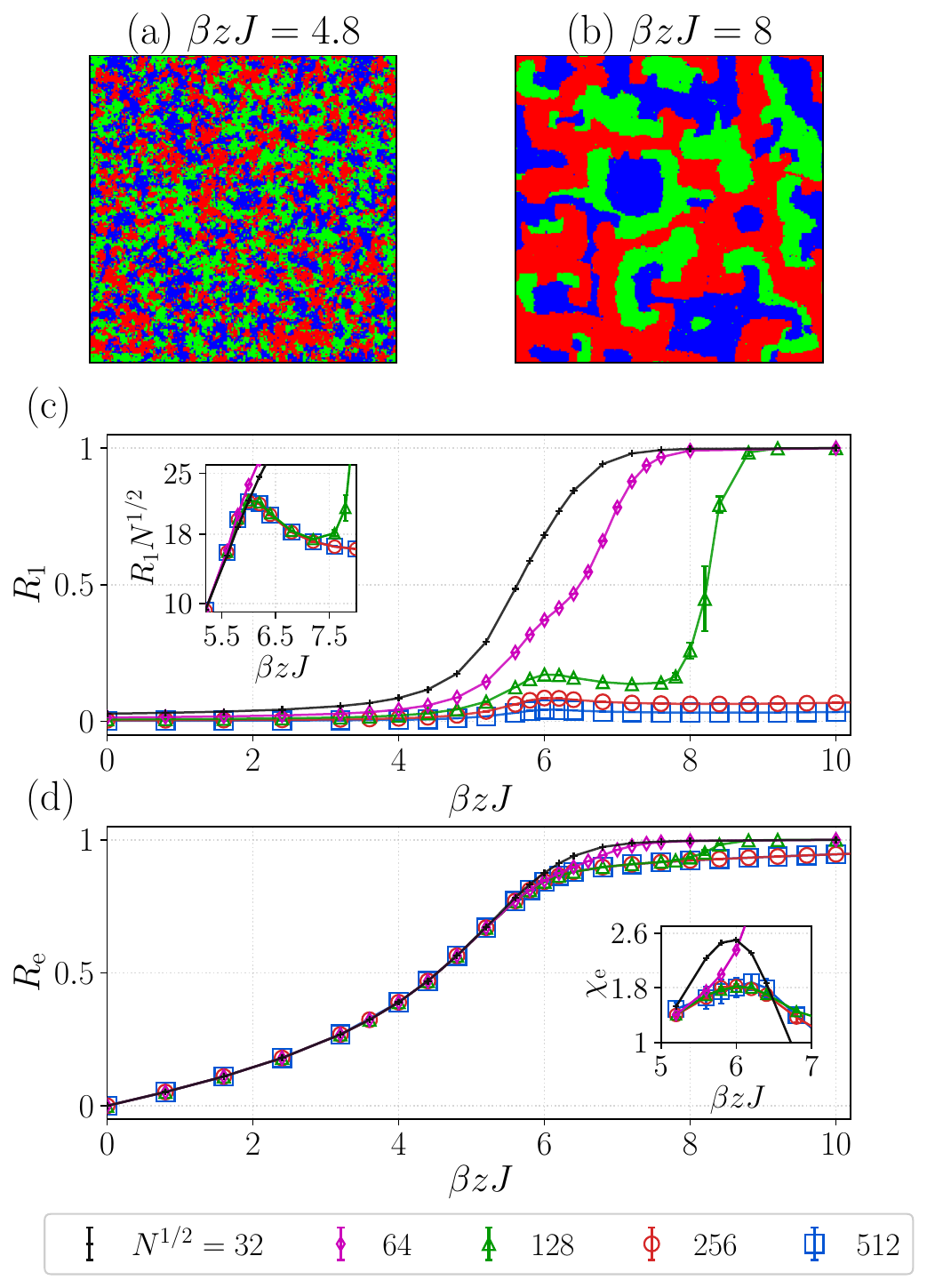}
    \caption{
        MC simulations of the three-state active Potts model on square lattices ($z=4$) at bias parameter $\zeta = 0.5$.
        (a), (b) Snapshots at system size $L/\sigma = 256$.
        (a) Disordered waves at $\beta z J = 4.8$.
        (b) Spiral waves at $\beta z J = 8$.
        Red, green, and blue squares represent the local states $s=0$, $1$, and $2$, respectively.
        (c), (d) Coupling-strength dependence of order parameters for various system sizes $L/\sigma = 32$, $64$, $128$, $256$, and $512$.
        (c) Single-state order parameter $R_1$.
        The inset shows the scaled order parameter $R_1 N^{1/2}$ near the transition.
        (d) Interfacial contact parameter $R_{\mathrm{e}}$.
        The inset shows the contact susceptibility $\chi_{\mathrm{e}}$.
        Samples trapped in the homogeneous cycling mode at $L/\sigma \geq 256$ are omitted from the ensemble averages for clarity.
    }
    \label{fig:mcz4}
\end{figure}

\begin{figure}[tbp]
    \centering
    \includegraphics[width=\columnwidth]{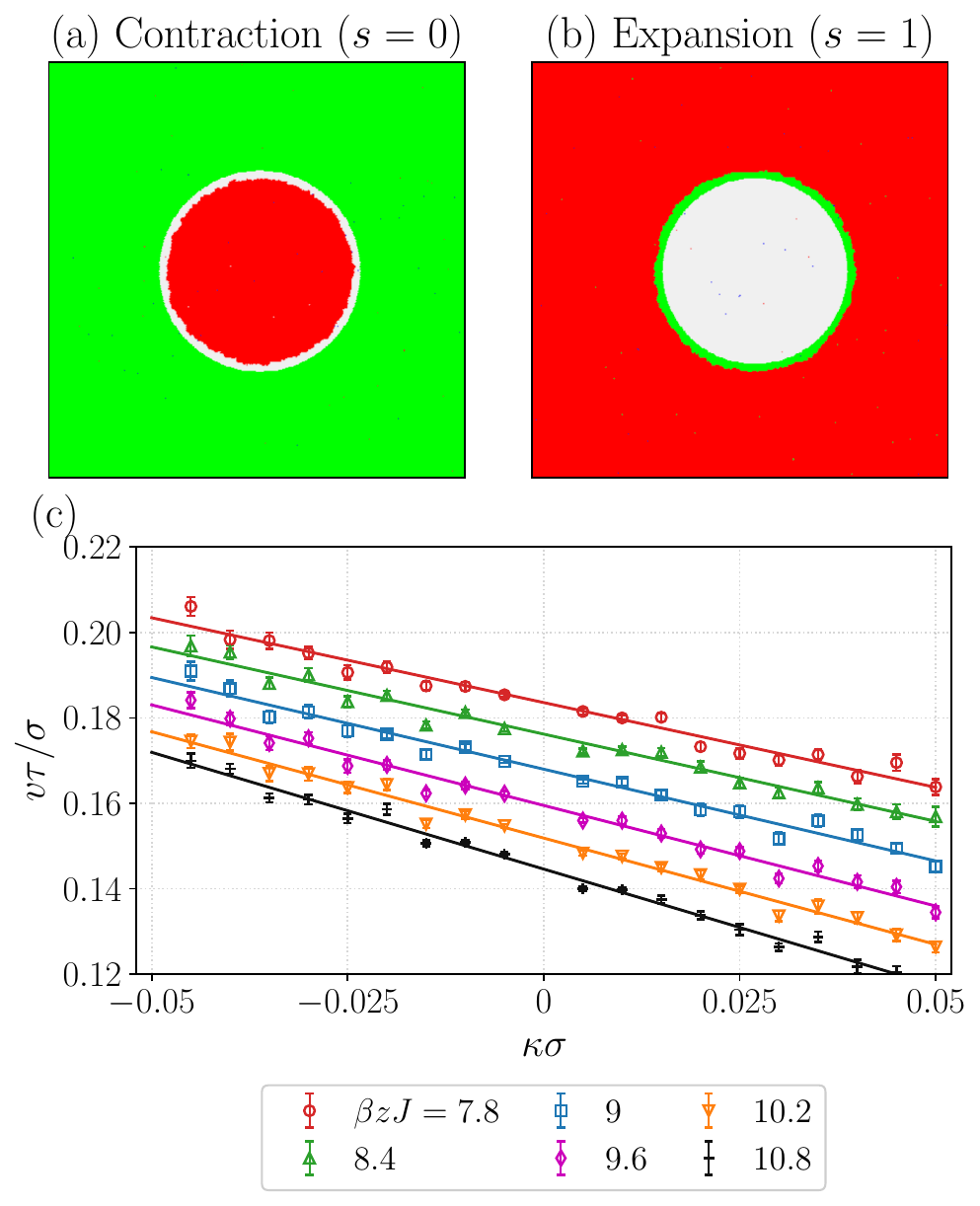}
    \caption{
        Curvature dependence of the dimensionless front velocity $V\tau/\sigma$ obtained by MC simulations of triangular lattices ($z=6$) at $\zeta = 0.5$ and $L/\sigma = 512$.
        (a)--(b) Superimposed snapshots of (a) contraction of an $s=0$ domain and (b) expansion of an $s=1$ domain at $R/\sigma = 100$.
        The red, green, and blue hexagons represent $s=0$, $1$, and $2$, respectively, in the final states.
        The white hexagons represent the sites within the initial domains whose final states are $s=1$.
        (c) Dependence of the dimensionless front velocity $V\tau/\sigma$ on the dimensionless curvature $\kappa\sigma$ at $\beta z J=7.8$, $8.4$, $9$, $9.6$, $10.2$, and $10.8$.
        The symbols represent the MC data.
        The solid lines show linear fits.
    }
    \label{fig:mccirc}
\end{figure}

\begin{figure}[tbp]
    \centering
    \includegraphics[width=\columnwidth]{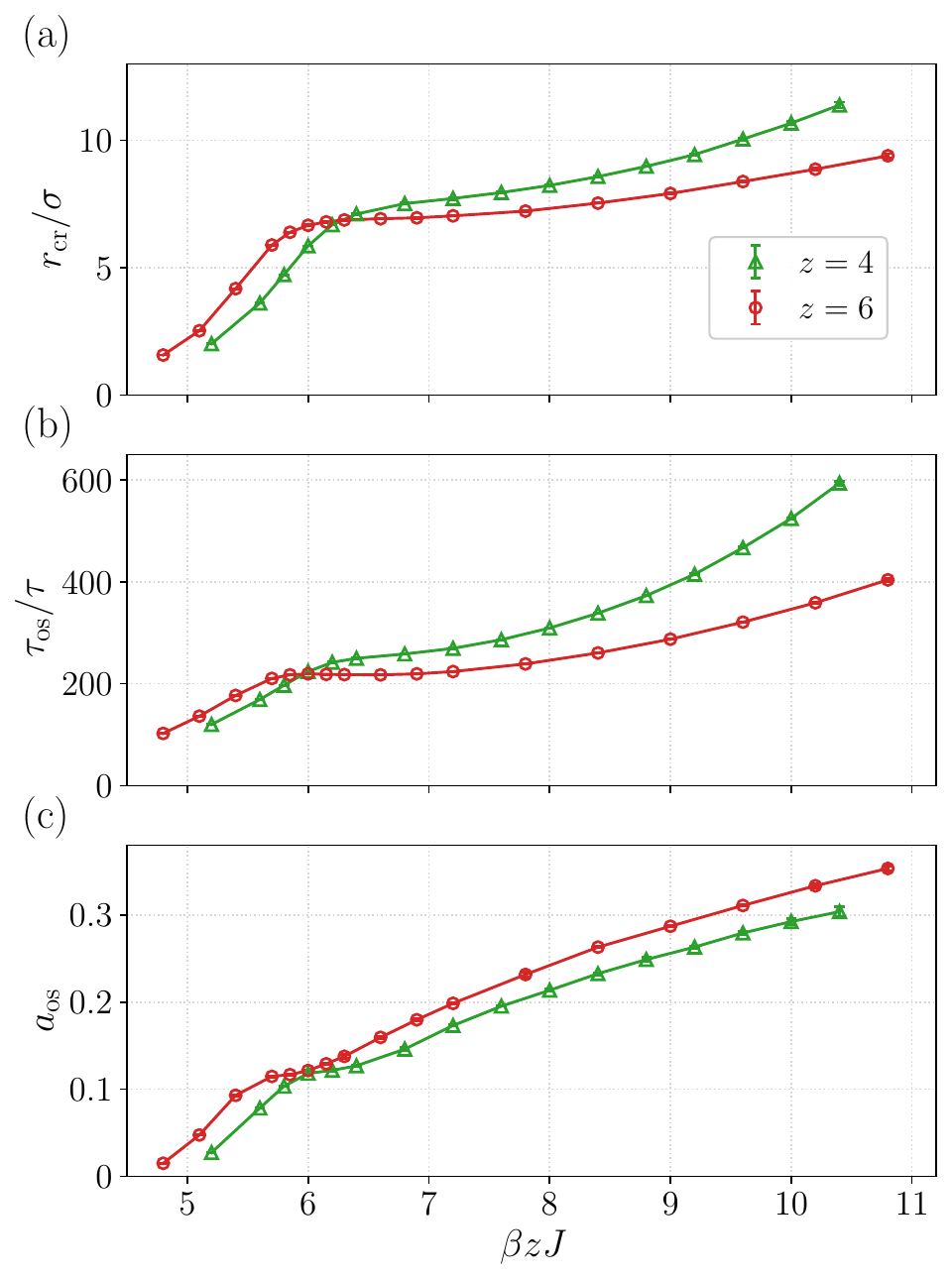}
    \caption{
        Spatial and temporal autocorrelation analysis in MC simulations at $\zeta = 0.5$ and $L/\sigma = 512$.
        (a) Dimensionless correlation length $r_{\mathrm{cr}}/\sigma$.
        (b)--(c) Dimensionless period $\tau_{\mathrm{os}}/\tau$ and amplitude $a_{\mathrm{os}}$ of the first peak in the temporal autocorrelation function $C_{\mathrm{t}}(t)$.
        The red circles and green triangles represent the data for $z=6$ and $z=4$, respectively.
    }
    \label{fig:mccrr}
\end{figure}

\section{Details of Monte Carlo Simulations}\label{sm:mc_details}
This section provides additional details on the estimation of the front velocity and the autocorrelation analyses for the lattice Monte Carlo (MC) simulations presented in Sec.~\ref{sec:mc_simulations}, corresponding to the data shown in Figs.~\ref{fig:mccirc} and \ref{fig:mccrr}.
To calculate the front velocity of circular domains, we measure the time interval $\tau_{R}$ over which the domain radius changes between $R + \Delta R$ and $R - \Delta R$ for the curvature $\kappa = 1/R$.
Hence, the front velocity is obtained as $V = 2\Delta R / \tau_{R}$.
Initially, for contraction, we prepare a circular domain of state $s = 0$ with radius $R + \Delta R$ embedded in a uniform background of state $s = 1$; for expansion, we prepare a circular domain of state $s = 1$ with radius $R - \Delta R$ in a background of state $s = 0$.
Here, we use $\Delta R = 4\sigma$.
Figures~\ref{fig:mccirc}(a) and \ref{fig:mccirc}(b) show superimposed snapshots of initial and final states for contraction and expansion, respectively.
The statistical errors are calculated from $20$ independent runs.
The obtained velocity is linearly dependent on the curvature $\kappa$ of the domain boundary as in the theoretical analysis and CMFE simulations, as shown in Fig.~\ref{fig:mccirc}(c).
The flat-interface propagation velocity $V_{\mathrm{flat}}$ and the effective diffusion coefficient $D_{\mathrm{eff}}$ in Eq.~\eqref{eq:eikonal} are estimated by linear regressions.

The front velocity $V$ is also estimated from the spatial and temporal autocorrelation functions.
The domain length is estimated from the spatial autocorrelation function $C_{\mathrm{r}}(\bm{r}) \coloneqq \langle\delta_{s_i,s_j}\delta(\bm{r}_{i,j}(t) - \bm{r})\rangle$~\cite{noguchi2026coarsening}.
The correlation is calculated along three principal axes [$(x,y) = (0, 1)$ and $(1/2, \pm \sqrt{3}/2)$] and along the $x$ and $y$ axes [$(x,y) = (1, 0)$ and $(0, 1)$] for triangular and square lattices, respectively.
The correlation length $r_{\mathrm{cr}}$ is defined as the half-relaxation length $C_{\mathrm{r}}(r_{\mathrm{cr}}) = (1 + \rho_{\mathrm{sat}})/2$ shown in Fig.~\ref{fig:mccrr}(a), where $\rho_{\mathrm{sat}} = 1/3$ is the saturation density.
The temporal autocorrelation function is defined as $C_{\mathrm{t}}(t) \coloneqq \langle \delta_{s_i(t+t_0),s_i(t_0)}\rangle$.
The oscillation amplitude $a_{\mathrm{os}}$ and period $\tau_{\mathrm{os}}$ are determined from the first peak of $C_{\mathrm{t}}(t)$ shown in Figs.~\ref{fig:mccrr}(b) and \ref{fig:mccrr}(c)~\cite{noguchi2026dynamics}: $\tau_{\mathrm{os}}$ is the time when the first peak occurs, and $a_{\mathrm{os}} = a_{\mathrm{pk}} - \rho_{\mathrm{sat}}$, where $a_{\mathrm{pk}}$ is the peak amplitude.
The front velocity is estimated as $V = \ell_{\mathrm{wave}}/\tau_{\mathrm{os}}$ and $\ell_{\mathrm{wave}} = 6r_{\mathrm{cr}}$, since the width of one domain is $2r_{\mathrm{cr}}$ and one wave consists of three domains.

\end{document}